\documentclass[authoryear,preprint,12pt]{elsarticle}

\usepackage{amssymb}
\usepackage{amsmath}
\usepackage{booktabs}
\usepackage{array}
\usepackage{tikz}
\usetikzlibrary{arrows.meta,positioning}

\journal{Coastal Engineering}

\begin{document}

\begin{frontmatter}



\title{Theoretical development of an operational wave-induced ice erosion model
       through laboratory experiments}

\author[ntnu]{Wenjun Lu\corref{cor1}}
\ead{wenjun.lu@ntnu.no}
\author[caen]{Behnam Ghadimi}
\author[caen]{Dominique Mouaze}
\author[caen]{Marianne Font}
\author[caen]{R\'{e}mi Lambert}
\author[npi]{Harvey Goodwin}
\author[ntnu]{Widar Weizhi Wang}
\author[ntnu]{Raed Lubbad}
\author[ntnu]{Sveinung L\o{}set}

\cortext[cor1]{Corresponding author}

\affiliation[ntnu]{organization={Department of Civil and Environmental Engineering,
                                 Norwegian University of Science and Technology (NTNU)},
            city={Trondheim},
            postcode={7491},
            country={Norway}}

\affiliation[caen]{organization={Laboratoire Morphodynamique Continentale et
                                 C\^{o}ti\`{e}re (M2C, CNRS UMR~6143),
                                 Universit\'{e} de Caen Normandie},
            addressline={24 rue des Tilleuls},
            city={Caen},
            postcode={14000},
            country={France}}

\affiliation[npi]{organization={Norwegian Polar Institute},
            addressline={Fram Centre, P.O. Box 6606 Stakkevollan},
            city={Troms\o{}},
            postcode={N-9296},
            country={Norway}}

\begin{abstract}
Wave-induced melting of vertical ice fronts is represented in several operational
iceberg and coastal-erosion models by the rough-wall parameterization of White
(1980), whose closure chain is incompletely documented and whose commonly used
compact expression is stated at the waterline. We reconstruct the formulation, specify the
rough-turbulent wave-friction closure using Jonsson's implicit relation and its
Lambert-W solution, and extend the model to a depth-resolved melt-rate profile under
linear wave kinematics. Because the horizontal and vertical orbital-velocity
components are linked, we use the horizontal component as a convenient
representative scale and introduce a dimensionless coefficient $\alpha$ for the
remaining closure uncertainty. The reconstruction gives a waterline coefficient of
$3.0\times10^{-4}$ with White's resultant-velocity definition and
$2.09\times10^{-4}$ for the reference choice $\alpha=1$; neither reproduces White's
published $1.46\times10^{-4}$ directly. Two monochromatic wave-flume experiments
with freshwater ice are then used to calibrate $\alpha$ from profiles below the
wave trough. The full Lambert-W friction coefficient is used in this calibration.
Best-fit values are 0.684 and 0.612 for periods of 1.54 and 0.87~s, respectively,
a relative difference of approximately 11\%. Their corresponding effective
waterline coefficients, $1.43\times10^{-4}$ and $1.28\times10^{-4}$, are close to
White's published value but do not constitute an independent validation. The fitted
profiles reproduce the observed depth dependence below the trough, while deviations
near the surface expose unresolved effects of intermittent submergence, local wave
impact, and uncertain thermal forcing.
\end{abstract}


\end{frontmatter}


\section{Introduction}
\label{sec:introduction}

Wave-induced melting at the face of floating or grounded ice bodies is a physically important process in polar and sub-polar coastal environments. Drifting icebergs, marine-terminating glaciers, marine and freshwater calving fronts, and tidally influenced permafrost bluffs are all subject to wave attack that enhances thermal melt through forced convection at the ice--water interface. Wave erosion is increasingly recognized as a primary mode of calving and ice-front retreat in ice-shelf and glacier-front settings \citep{sartore2025,rohl2006}. Quantifying this thermal contribution to ice loss is a necessary component of predictive models for iceberg drift and decay \citep{kubat2007,gladstone2001,eik2009}, mass balance of calving glaciers, and the stability of ice-rich permafrost coastlines \citep{barnhart2014}. Despite this wide relevance, a central operational parameterization still traces back to the foundational study of \citet{white1980}.

\citet{white1980} derived a compact parameterization for the wave-induced melt rate at the waterline of a vertical ice body by coupling linear wave theory with oscillatory boundary-layer heat transfer. In its most widely used form, the formula expresses the waterline melt rate as a simple function of wave height, period, surface roughness, and water--ice temperature difference (Eq.~\eqref{eq:White_waterline}). The formula has since influenced iceberg deterioration models \citep{kubat2007,keghouche2010,eltahan1987,eik2009}, wave-erosion estimates for ice shelves \citep{gladstone2001}, and estimates of wave-forced erosion of permafrost coastlines \citep{barnhart2014}. The problem has also been studied for freely drifting icebergs, where body motion modifies the wave field \citep{forouzi2025a,forouzi2025b}. For a fixed vertical wall under non-breaking, near-deep-water monochromatic waves, \citet{wolterman2026} derived and experimentally evaluated a depth-dependent melt law based on wave-averaged boundary-layer streaming and thermal diffusion. The present study addresses a complementary problem by reconstructing the rough-turbulent Stanton--friction framework used in White-derived operational models, specifying its Jonsson closure, and evaluating its finite-depth profile against two wave-flume conditions.

Despite its widespread use, the \citet{white1980} formulation contains several unresolved theoretical and experimental issues, documented in Section~\ref{sec:white_framework}. They include an incompletely documented derivation chain and unspecified friction coefficient, the lack of an explicit finite-depth form in White's original presentation, ambiguity in the characteristic velocity scale, and limited laboratory evaluation of the rough-wall form.

This study addresses these issues through theoretical reconstruction and controlled laboratory experimentation. The specific contributions are as follows. (i)~The \citet{white1980} framework is reconstructed using the rough-turbulent oscillatory boundary-layer relations of \citet{jonsson1966} to specify the friction coefficient. (ii)~White's waterline expression is extended to a depth-resolved form that accounts for the variation of wave kinematics with depth under linear wave theory. (iii)~A velocity scaling parameter $\alpha$ is introduced for the unresolved velocity closure and calibrated against two monochromatic wave-flume experiments. These tests provide an experimental benchmark for the reconstructed rough-wall formulation, not an independent validation of a coefficient fitted to the same data.

The paper is structured as follows. Section~\ref{sec:theoretical_background} presents the White~(1980) framework, documents the main unresolved issues, and derives the full-depth formulation. Section~\ref{sec:methods} describes the laboratory experimental setup. Section~\ref{sec:results} reports the experimental observations and the theoretical comparison, including the calibrated $\alpha$ values. Section~\ref{sec:discussion} discusses the physical interpretation of the calibration results, the limitations of the framework, and the splash-zone behaviour. Section~\ref{sec:conclusions} summarises the main findings.

\section{Theory}
\label{sec:theoretical_background}

\subsection{White (1980) and unresolved issues}
\label{sec:white_framework}

\begin{figure}[htbp]
    \centering
    \begin{tikzpicture}[x=1.0cm,y=1.15cm,font=\small]
        \fill[blue!24] (-6,-3.2) rectangle (0,0);
        \fill[black!9] (0,-3.2) rectangle (5.8,0.85);
        \draw[thick] (-6,-3.2)--(5.8,-3.2);
        \draw[very thick] (0,-3.2)--(0,0.85);
        \draw[thick] (0,0.85)--(5.8,0.85);

        \draw[thick,blue!65!black] (-6,0)--(0,0);
        \draw[thick,blue!65!black,smooth,domain=-5.7:-3.7,samples=80]
            plot (\x,{0.18*sin(360*(\x+5.7)/1.25)});
        \draw[-{Latex[length=5pt]},thick,blue!65!black] (-5.6,-0.55)--(-4.2,-0.55);

        \draw[-{Latex[length=6pt]},ultra thick] (0,0)--(5.25,0)
            node[below left=2pt] {$x$};
        \draw[-{Latex[length=6pt]},ultra thick] (0,-3.2)--(0,2.0)
            node[left=2pt] {$z$};
        \fill (0,0) circle (1.5pt);

        \draw[<->,thick,gray!70!black] (-0.65,-3.2)--(-0.65,0);
        \node[rotate=90,fill=blue!24,inner sep=2pt] at (-0.65,-1.6) {$d$};

        \node[blue!55!black] at (-3.0,-1.55) {Water};
        \node at (3.0,0.42) {Fixed ice body};
        \node[above,align=center] at (-4.7,0.35) {Incoming wave\\($H,T$)};
        \node[above left] at (-0.08,0.03) {MSL};
        \node[rotate=90,anchor=south] at (0.08,-1.6) {ice--water interface};
        \node[below] at (-3.0,-3.2) {seabed};
    \end{tikzpicture}
    \caption{Problem configuration. An incoming monochromatic wave of height $H$ and period $T$ propagates toward a fixed, semi-submerged vertical ice body. The origin is at the initial ice--water interface at mean sea level (MSL), with $z$ positive upward and $x$ positive into the ice. The body extends from the seabed ($z=-d$) to above the wave crest, and its face retreats in the positive $x$-direction at the depth-dependent melt rate $V_m(z)$.}
    \label{fig:problem_setup}
\end{figure}
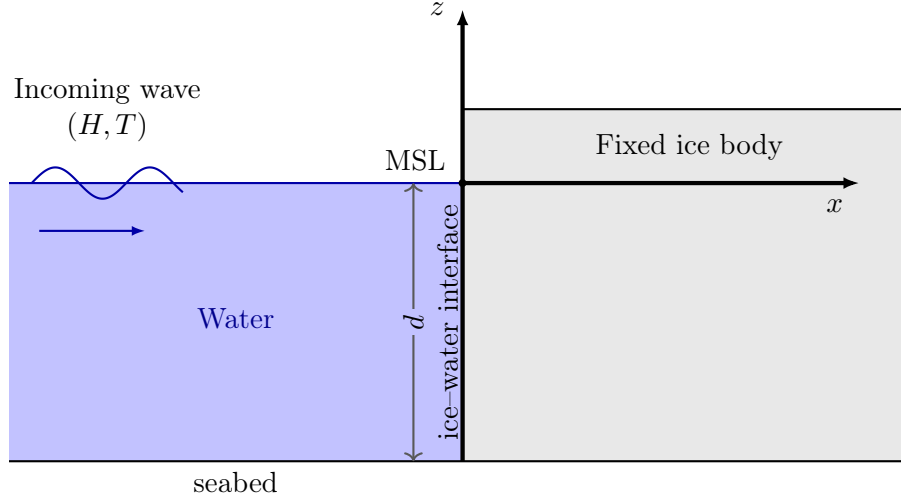

The physical configuration is illustrated in Fig.~\ref{fig:problem_setup}: a semi-submerged ice body is fixed, with its vertical face perpendicular to the direction of wave propagation. Coordinates are centred at the initial ice--water interface at mean sea level (MSL), with $z$ positive upward and $x$ positive into the ice. The formulation assumes (i)~linear (Airy) wave theory, (ii)~no depth-limited breaking in the incident wave field, (iii)~a fixed, passive ice boundary, and (iv)~the local wave height at the ice face as the relevant forcing parameter.

The wave-induced melting parameterization of \citet{white1980}, in its most widely applied form, expresses the melt rate at the waterline as
\begin{equation}
\frac{V_m T}{H}
=
0.000146\left(\frac{k_s}{H}\right)^{0.2}\Delta T,
\label{eq:White_waterline}
\end{equation}
where $V_m$ [\,m\,s$^{-1}$] is the melt (recession) rate, $T$ [\,s] is the wave period, $H$ [\,m] is the wave height, $k_s$ [\,m] is the surface roughness length, and $\Delta T$ [\,K] is the water--ice temperature difference. Related wave-erosion parameterizations have been used for over four decades in iceberg deterioration models \citep{kubat2007,keghouche2010,eltahan1987,eik2009}, ice-shelf erosion estimates \citep{gladstone2001}, and permafrost coastal-erosion estimates \citep{barnhart2014}. The intermediate steps between White's broader framework and this compact rough-wall expression are not documented in the report.

A careful reconstruction of that derivation reveals four issues that are directly relevant to the present paper.

\textbf{Friction coefficient unspecified.}
\label{sec:missing_friction}
No explicit expression for the rough-turbulent wave friction coefficient $C_f$ is given in White (1980). Since $C_f$ controls the Stanton-number scaling, the missing closure obscures how the numerical coefficient in the waterline formula should be reproduced or adapted outside the original parameter range.

\textbf{Depth profile absent.}
\label{sec:missing_depth}
White (1980) provides only the waterline melt rate. This is insufficient when the objective is to predict the full eroded ice-front shape, because both orbital velocity and contact time vary with elevation.

\textbf{Velocity scaling ambiguous.}
\label{sec:velocity_scaling}
White defines the characteristic velocity as the resultant orbital magnitude $V = \sqrt{u_m^2 + w_m^2}$, combining horizontal and vertical components without a unique physical closure. Because the orbital-velocity components are related through the same linear wave field, the choice of a representative velocity scale is partly conventional. The present study adopts the horizontal orbital velocity $u_m$ as a simple and reproducible reference scale, writes $V = \alpha\,u_m$, and calibrates $\alpha$ against laboratory data.

\textbf{Coefficient discrepancy and limited validation.}
\label{sec:mismatch}
Reconstructing the waterline formula from the documented closure chain does not unambiguously lead to White's published coefficient $1.46 \times 10^{-4}$. In addition, White's own laboratory experiments validated only the smooth-wall version of his formula; the rough-wall coefficient was not directly tested against experimental data in the original study. These issues make it difficult to separate derivation uncertainty from empirical calibration.

\subsection{Full-depth melt-rate formulation}
\label{sec:full_depth_derivation}

\begin{figure}[htbp]
\centering
\begin{tikzpicture}[
  node distance=0.38cm,
  box/.style={
    rectangle, draw=black!65, rounded corners=3pt,
    minimum width=10cm, text width=9.8cm,
    minimum height=0.68cm, align=center, font=\small,
    fill=gray!6
  },
  outbox/.style={box, fill=blue!10, draw=blue!55!black},
  arr/.style={-{Latex[length=4pt,width=4pt]}, thick, black!55}
]
  \node[box] (A) {Wave input: $H$, $T$, $d$ \quad$\longrightarrow$\quad
                  wavenumber $k$ via $\omega^2 = gk\tanh(kd)$};
  \node[box, below=of A] (B)
    {Orbital velocity \& excursion [Eqs.~\ref{eq:um}--\ref{eq:a1m}]:\quad
     $u_m(z)$,\; $a_{1m}(z)$};
  \node[box, below=of B] (C)
    {Amplitude Reynolds number [Eq.~\ref{eq:Rea_depth}]:\quad
     $\mathrm{Re}_a(z) = u_m(z)\,a_{1m}(z)/\nu$};
  \node[box, below=of C] (D)
    {Friction coefficient [\ref{app:Cf_derivation}, Eq.~\ref{eq:Cf_LambertW}]:\quad
     $C_f(a_{1m}/k_s)$ \enspace \textit{lab}: Lambert-W;\enspace \textit{field}: power-law};
  \node[box, below=of D] (E)
    {Stanton number [\ref{app:St_algebra}]:\quad
     $\mathrm{St}(z)$ \enspace \textit{compact power-law form}: Eq.~\ref{eq:StWhite}};
  \node[box, below=of E] (F)
    {Heat flux [Eq.~\ref{eq:q_w}]:\quad
     $q_w = \mathrm{St}\,\rho_w\,(\alpha\,u_m)\,c_p\,\Delta T$\quad
     [$\alpha$ calibrated, \S\ref{sec:alpha_calibration}]};
  \node[outbox, below=of F] (G)
    {\textbf{Depth-resolved melt rate [compact form: Eq.~\ref{eq:Vm_full}]:}\quad
     $V_m(z) = q_w\,/\,(\rho_i\,\Gamma)$};

  \draw[arr] (A)--(B); \draw[arr] (B)--(C); \draw[arr] (C)--(D);
  \draw[arr] (D)--(E); \draw[arr] (E)--(F); \draw[arr] (F)--(G);
\end{tikzpicture}
\caption{Derivation chain from wave-input parameters to the depth-resolved melt rate $V_m(z)$.
         Equation~\protect\ref{eq:Vm_full} gives the compact form obtained with the power-law
         friction coefficient.
         The sole empirically determined quantity is the velocity scaling parameter $\alpha$;
         all other quantities follow from wave kinematics, Jonsson's (1966) boundary-layer
         theory, and the thermodynamic constants listed in Table~\protect\ref{tab:ice_properties}.}
\label{fig:derivation_chain}
\end{figure}
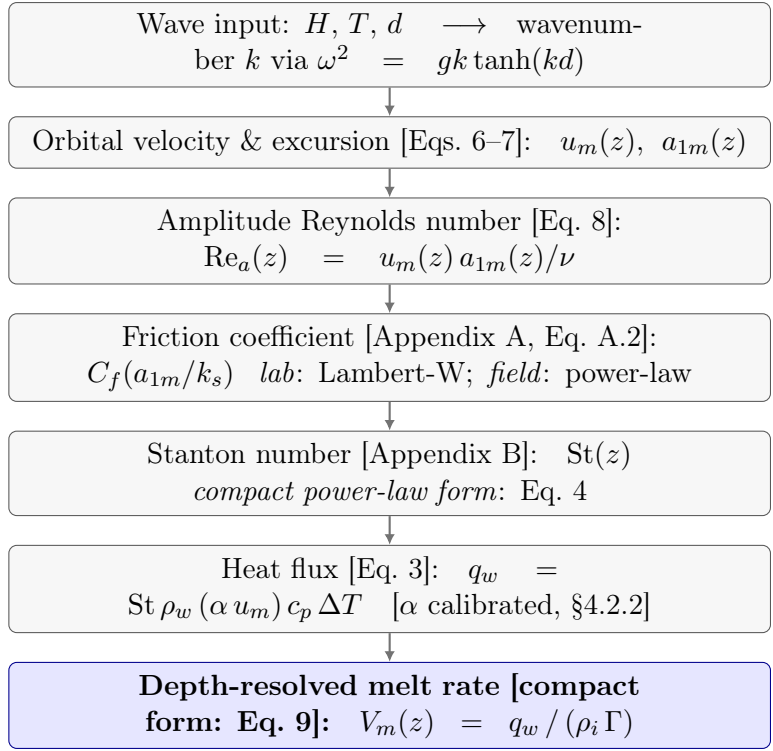

Figure~\ref{fig:derivation_chain} summarises the derivation chain from wave-input parameters to the depth-resolved melt-rate formulation; the sub-sections below detail each step.

\subsubsection{Energy balance and Stanton heat flux}

White (1980) formulates melting as a pure heat-budget problem at the ice--water interface: heat delivered from the water is instantaneously expended in melting ice, with meltwater efficiently removed by wave-driven turbulence. Internal ice conduction is neglected; ice conductivity is approximately $2.1$\,W\,m$^{-1}$\,K$^{-1}$. The latent heat of fusion ($\Gamma = 3.34 \times 10^5$\,J\,kg$^{-1}$) dominates the sensible-heat budget by an order of magnitude. The melt rate is therefore determined solely by the interfacial heat flux $q_w$,
\begin{equation}
V_m = \frac{q_w}{\rho_i \Gamma},
\label{eq:Vm}
\end{equation}
where $\rho_i$ [kg\,m$^{-3}$] is ice density and $q_w$ [W\,m$^{-2}$] is the interfacial heat flux. White parameterizes the heat flux via the Stanton-number formulation
\begin{equation}
q_w = \mathrm{St}\,\rho_w\,V\,c_p\,\Delta T,
\label{eq:q_w}
\end{equation}
where $\mathrm{St}$ [-] is the Stanton number, $\rho_w$ [kg\,m$^{-3}$] is water density, $V$ [m\,s$^{-1}$] is a characteristic velocity scale, and $c_p$ [J\,kg$^{-1}$\,K$^{-1}$] is the specific heat capacity of water.

\subsubsection{Compact Stanton-number result}

\citet{white1980} relates the Stanton number to wave-induced boundary friction via the oscillatory boundary-layer closure of \citet{jonsson1966}, treating the ice face as hydraulically equivalent to a wave-bottom boundary. When the rough-wall friction coefficient is represented by the power-law approximation in Eq.~\eqref{eq:Cf}, the algebraic reduction from the Stanton--friction scaling relation to the compact closed form is detailed in \ref{app:St_algebra}. The result is
\begin{equation}
\mathrm{St}
=
\frac{0.0557\left(\dfrac{k_s}{a_{1m}}\right)^{0.2}}
{\mathrm{Re}_a^{0.1} + \dfrac{3.84}{\sqrt{2}}\left(\mathrm{Pr}^{0.68}-1\right)},
\label{eq:StWhite}
\end{equation}
where $\mathrm{Re}_a = u_m a_{1m}/\nu$ [-] is the amplitude Reynolds number, $u_m$ [m\,s$^{-1}$] is the horizontal orbital velocity amplitude, $a_{1m}$ [m] is the orbital excursion amplitude, $\nu$ [m$^2$\,s$^{-1}$] is kinematic viscosity, and $\mathrm{Pr}$ [-] is the Prandtl number.

\subsubsection{Modified velocity closure}

White's formulation does not uniquely specify the characteristic velocity entering the heat-transfer closure. The horizontal and vertical orbital-velocity components are linked through the same linear wave field. We therefore adopt the horizontal orbital velocity $u_m$ as a simple representative scale and absorb the remaining closure uncertainty into a dimensionless factor:
\begin{equation}
V = \alpha\,u_m,
\label{eq:V_alpha}
\end{equation}
where $\alpha$ [-] is a calibration coefficient that absorbs velocity scaling uncertainty. The choice $\alpha = 1$ (horizontal component only) is adopted as the theoretical reference; White's original resultant-velocity definition corresponds to $\alpha = \sqrt{2}$ at the deep-water waterline.

\subsubsection{Wave-kinematic closure}

Under linear wave theory (Airy), the horizontal orbital velocity amplitude and orbital excursion amplitude are
\begin{equation}
u_m(z) = \frac{\pi H}{T}\cos\phi\,
\frac{\cosh\!\left[k(z+d)\right]}{\sinh(kd)},
\label{eq:um}
\end{equation}
\begin{equation}
a_{1m}(z) = \frac{H}{2}\,
\frac{\cosh\!\left[k(z+d)\right]}{\sinh(kd)},
\label{eq:a1m}
\end{equation}
where $k$ [m$^{-1}$] is the wavenumber, $\phi$ [-] is the wave incidence angle, and $d$ [m] is water depth. The wavenumber satisfies the linear dispersion relation $\omega^2 = gk\tanh(kd)$ with $\omega = 2\pi/T$ [s$^{-1}$].

\subsubsection{Full-depth melt-rate expression}
\label{sec:theoretical_rederivation}

Combining Eqs.~\eqref{eq:Vm}--\eqref{eq:a1m} gives the depth-dependent amplitude Reynolds number,
\begin{equation}
\mathrm{Re}_a(z) = \frac{u_m(z)\,a_{1m}(z)}{\nu}
=
\frac{\pi\,\cos\phi\,H^2}{2T\nu}
\left[\frac{\cosh\!\left[k(z+d)\right]}{\sinh(kd)}\right]^{\!2},
\label{eq:Rea_depth}
\end{equation}
and the compact power-law form of the depth-resolved melt-rate expression,
\begin{equation}
\boxed{
\frac{T}{H}V_m
=
0.0027\,\alpha
\cos\phi
\left(\frac{\cosh\!\left[k(z+d)\right]}{\sinh(kd)}\right)^{\!0.8}
\frac{\left(\dfrac{k_s}{H}\right)^{0.2}}
{\mathrm{Re}_a(z)^{0.1}
+
\dfrac{3.84}{\sqrt{2}}\left(\mathrm{Pr}^{0.68}-1\right)}
\;\Delta T,
}
\label{eq:Vm_full}
\end{equation}
where the prefactor $0.0027~\mathrm{K}^{-1} = (\rho_w c_p/\rho_i\Gamma)\,\pi \times 0.0557 \times 2^{0.2}$ absorbs thermodynamic constants, the Stanton-number coefficients, and the factor $2^{0.2}$ from $a_{1m}(0) = H/2$ at the waterline. The units arise from $(\rho_w c_p)/(\rho_i\Gamma)$, while $\alpha$ is dimensionless; thus the product of the prefactor with $\Delta T$ [K] is dimensionless, consistent with the left-hand side $TV_m/H$. This compact expression follows from the power-law friction coefficient
\begin{equation}
C_f = 0.138\left(\frac{k_s}{a_{1m}}\right)^{0.4},
\label{eq:Cf}
\end{equation}
which reproduces the Lambert-W solution to within 7\% over $a_{1m}/k_s \approx 30$--$300$. For the laboratory evaluation, where $a_{1m}/k_s \approx 4$--$6$, the power-law approximation is outside this accuracy range; the Stanton relation is therefore evaluated before the power-law substitution, using the Lambert-W friction coefficient
\begin{equation}
C_f = \left(\frac{\ln 10}{4\,W\!\left(1.916\,\dfrac{a_{1m}}{k_s}\right)}\right)^{\!2},
\tag{\ref{eq:Cf_LambertW}}
\end{equation}
where $W(\cdot)$ is the principal branch of the Lambert-W function. The derivation of this closed-form expression from Jonsson's implicit relation is given in \ref{app:Cf_derivation}.

\subsubsection{Reduction to White's waterline formula}
\label{sec:reduction}

Evaluating Eq.~\eqref{eq:Vm_full} at $z=0$, $kd\to\infty$, $\phi=0$, and approximating the denominator by the dominant Prandtl-number term (\ref{app:St_algebra}) recovers White's waterline form with an effective coefficient $\alpha C'$. For $\alpha=1$ this gives $C'_{\rm ours} \approx 2.09\times10^{-4}$; for $\alpha=\sqrt{2}$ it gives $C' \approx 3.0\times10^{-4}$, recovering the value derived independently in \ref{app:discrepancy}. The complete reduction and coefficient evaluation are given in \ref{app:discrepancy}. Following White's documented logic does not reproduce his published coefficient $1.46\times10^{-4}$; the discrepancy therefore reflects an unresolved step in the original reduction rather than an algebraic consequence of the stated assumptions.

\begin{table}[htbp]
\centering
\caption{Summary of the three reference values of the waterline melt coefficient.
         The present study adopts $V = \alpha\,u_m$ with $\alpha = 1$ as reference.}
\label{tab:coeff_landscape}
\begin{tabular}{lcc}
\hline
Velocity definition & Derived $C'$ & White's published \\
\hline
$V = \sqrt{u_m^2+w_m^2}$ (White's) & $3.0\times10^{-4}$ & $1.46\times10^{-4}$ \\
$V = u_m$ (present baseline, $\alpha=1$) & $2.09\times10^{-4}$ & --- \\
\hline
\end{tabular}
\end{table}

\section{Laboratory Experiments}
\label{sec:methods}

\subsection{Laboratory experiment}
\label{sec:lab_experiment}

A laboratory experiment was conducted to generate benchmark data for evaluating and reparameterizing the velocity scale in the White (1980) formulation. The experiment was designed to expose a freshwater ice block to regular monochromatic waves in a controlled flume environment and to track the evolution of the ice front profile over time. Two valid test runs were completed, representing qualitatively distinct wave regimes that will be referred to throughout as Test~1 and Test~2.

\subsubsection{Test setup}
\label{sec:test_setup}

\begin{figure}[htbp]
    \centering
    \includegraphics[width=\textwidth]{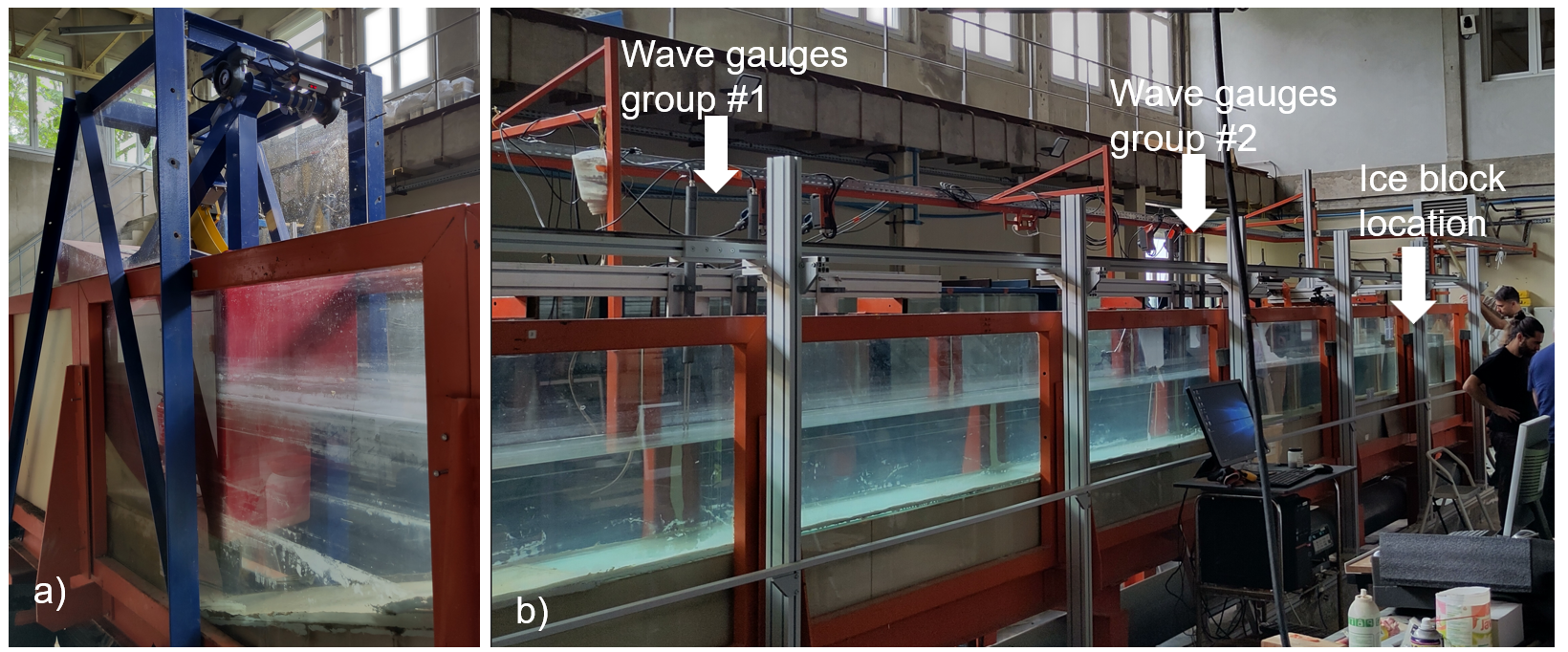}
    \caption{Wave flume overview. (a)~Wave paddle at the seaward end of the flume, where monochromatic waves are generated. (b)~Wide-angle view of the full flume showing the locations of Wave Gauges Group~1 (near the wave paddle), Wave Gauges Group~2 (near the ice block), and the ice block at the landward end.}
    \label{fig:flume_overview}
\end{figure}

Experiments were carried out in a wave flume at the Laboratoire Morphodynamique Continentale et C\^{o}ti\`{e}re (M2C, CNRS UMR~6143), Universit\'{e} de Caen Normandie, Caen, France. The flume is equipped with transparent glass side walls, enabling direct side-view optical observation of the ice front throughout each test. The water depth was maintained at $d = 0.31$~m for all runs. The ice block was placed at the landward end of the flume, with its vertical face oriented perpendicular to the direction of wave propagation. An overview of the flume configuration is shown in Fig.~\ref{fig:flume_overview}.

\begin{figure}[htbp]
    \centering
    \includegraphics[width=\textwidth]{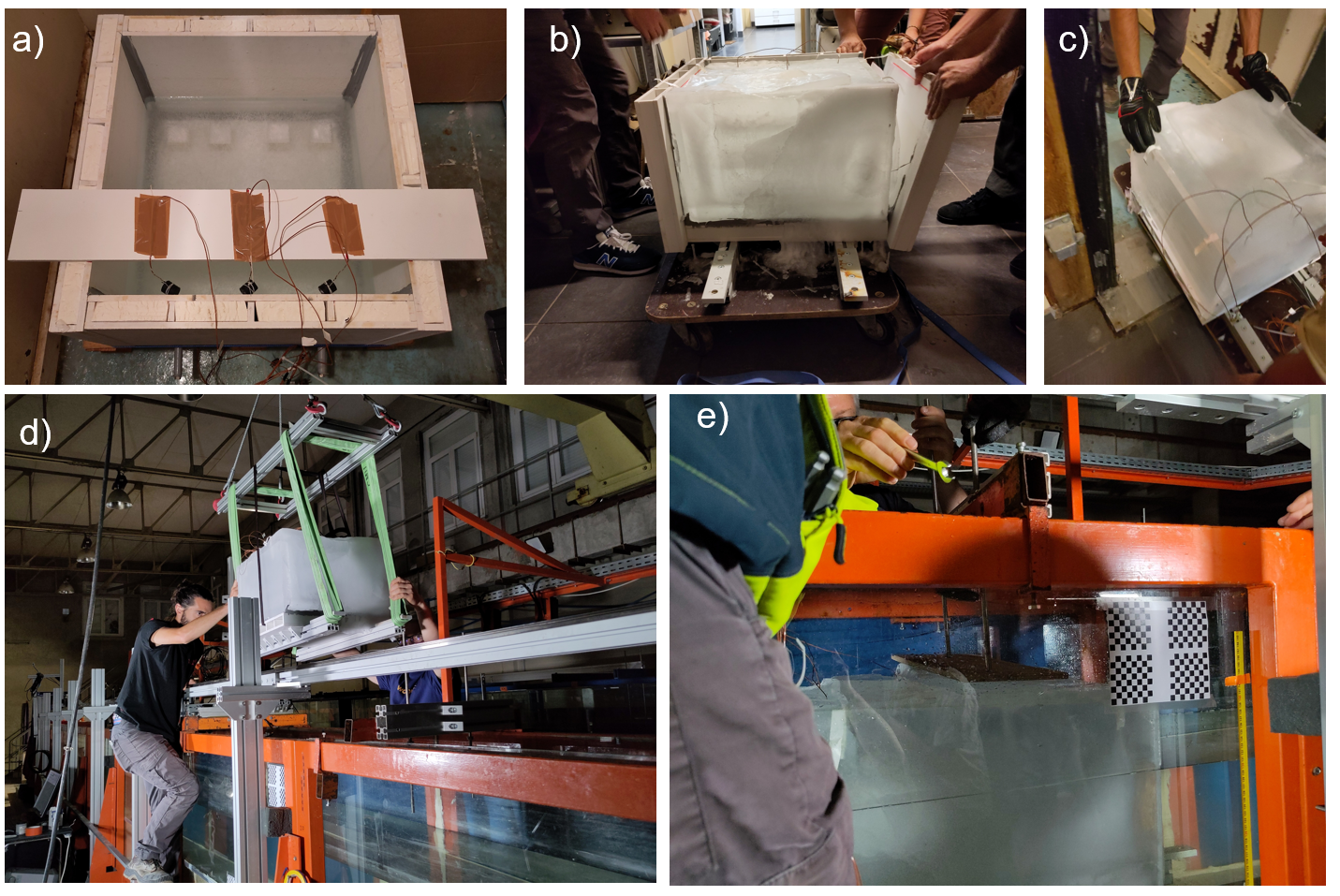}
    \caption{Overview of the experimental procedure. (a)~Ice block in the polystyrene mould during freezing in the cold room; iButton temperature sensors are visible as dark spots embedded within the ice. (b)~Demoulded ice block ready for transport. (c)~Transfer of the ice block from the cold room ($-20\,^\circ$C) to the wave flume. (d)~Positioning of the ice block at the landward end of the wave flume. (e)~Installed configuration: the steel frame and plate anchor the block vertically while leaving the front face free to melt.}
    \label{fig:procedure}
\end{figure}

The overall experimental procedure is illustrated in Fig.~\ref{fig:procedure}. Ice blocks were fabricated by filling insulated polystyrene moulds with freshwater in successive layers and storing them in a cold room at approximately $-20\,^\circ$C for several days until fully frozen (panel~a). During filling, iButton temperature sensors were embedded at prescribed positions within the ice (panel~a). After freezing was complete, the mould was removed (panel~b) and the ice block was transferred from the cold room ($-20\,^\circ$C) to the wave flume (panel~c) and positioned at the reflective end (panel~d). Once in position, a rigid steel frame with two adjustment screws was anchored to the flume walls and a steel plate was pressed against the top surface of the ice to provide vertical restraint, preventing the block from floating while leaving the front face free to erode (panel~e).

\begin{figure}[htbp]
    \centering
    \includegraphics[width=\textwidth]{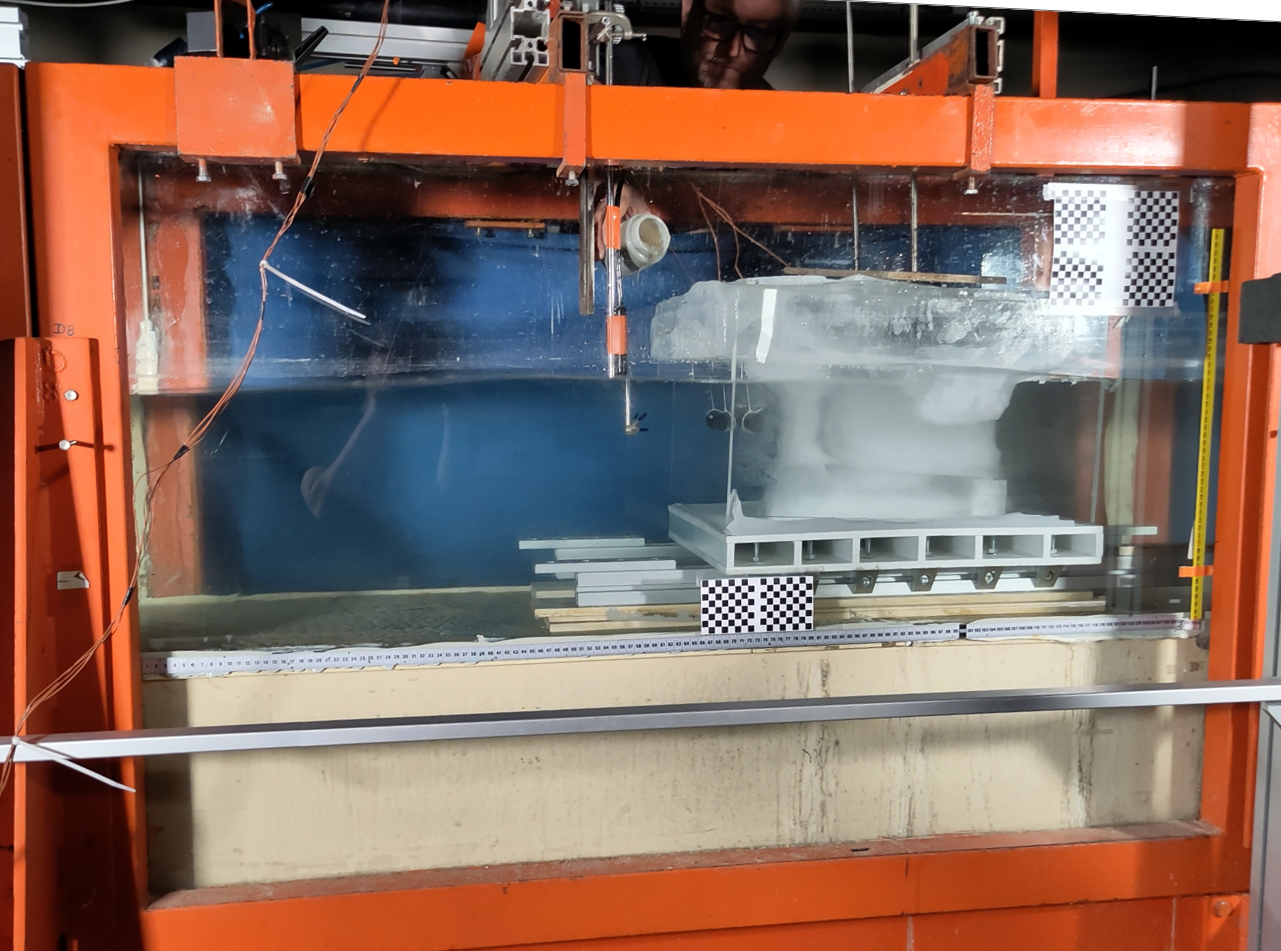}
    \caption{Close-up view of the test configuration at the landward end of the wave flume. The orange rectangular outline is the outer flume frame and was used as the geometric reference for image rectification. The white horizontal ruler and yellow vertical ruler provided independent pixel-to-physical scale calibration in the two image directions.}
    \label{fig:flume_setup}
\end{figure}

The orange rectangular outline visible in the images is the outer frame of the wave flume itself. Because this frame is rectangular, its four corners were used as the geometric reference for image rectification in the subsequent image analysis (see Section~\ref{sec:data_processing} and Fig.~\ref{fig:flume_setup}). Physical scaling was then obtained from two rulers visible in the camera view: a white ruler along the horizontal direction and a yellow ruler along the vertical direction. These provided independent horizontal and vertical pixel-to-physical scale factors. A checkerboard calibration panel was also present in the flume; however, due to the small size of individual squares relative to the image resolution, the flume-frame corners were adopted as the rectification reference instead.

\subsubsection{Instrumentation}
\label{sec:instrumentation}

Wave conditions were monitored using two groups of three resistance-type wave gauges positioned along the flume centreline at specified separations (Fig.~\ref{fig:flume_overview}). Each three-gauge array enables decomposition of the measured free-surface elevation signal into incident and reflected components using the least-squares method of \citet{mansardfunke1980}.
A digital camera was mounted externally on a fixed tripod aligned with the centre of the glass wall, recording the ice face in continuous video throughout each test. Individual frames were subsequently extracted from the video at uniform one-minute intervals, yielding a time series of ice-front profiles at 60-second resolution.

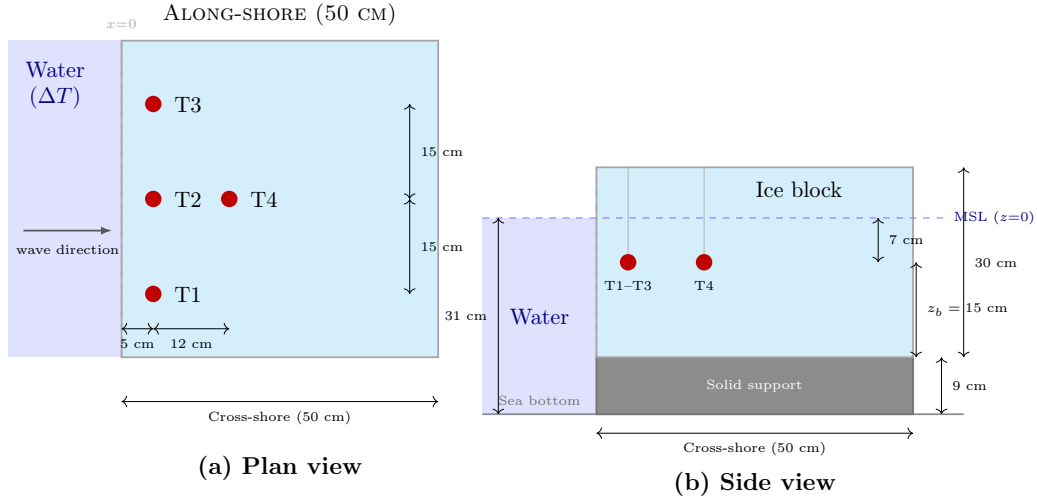
\begin{figure}[htbp]
\centering
\resizebox{\textwidth}{!}{%
\begin{tikzpicture}[font=\small, node distance=0pt]

\begin{scope}[xshift=0cm, yshift=0cm]

  \fill[blue!12] (-1.8,0) rectangle (0,5);
  \node[align=center, font=\footnotesize, text=blue!50!black] at (-1.05,4.3)
    {Water\\($\Delta T$)};

  \fill[cyan!15] (0,0) rectangle (5,5);
  \draw[thick, black!35] (0,0) rectangle (5,5);

  \node[above=0.1cm, font=\footnotesize] at (2.5,5) {\textsc{Along-shore (50 cm)}};

  \draw[thick, dashed, gray!65] (0,0)--(0,5);
  \node[above, font=\tiny, gray!65] at (0,5.05) {$x{=}0$};

  \draw[-{Latex[length=5pt]}, thick, black!65] (-1.55,2.0)--(-0.15,2.0);
  \node[below, font=\tiny] at (-0.85,1.92) {wave direction};

  \foreach \i/\y in {1/1.0, 2/2.5, 3/4.0} {
    \fill[red!75!black] (0.5, \y) circle (0.13);
    \node[right=0.06cm, font=\footnotesize] at (0.63, \y) {T\i};
  }
  \fill[red!75!black] (1.7,2.5) circle (0.13);
  \node[right=0.06cm, font=\footnotesize] at (1.83,2.5) {T4};

  \draw[<->] (0,0.45)--(0.5,0.45);
  \node[below, font=\tiny] at (0.25,0.43) {5 cm};
  \draw[<->] (0.5,0.45)--(1.7,0.45);
  \node[below, font=\tiny] at (1.1,0.43) {12 cm};
  \draw[<->] (4.55,1.0)--(4.55,2.5);
  \draw[<->] (4.55,2.5)--(4.55,4.0);
  \node[right, font=\tiny] at (4.58,1.75) {15 cm};
  \node[right, font=\tiny] at (4.58,3.25) {15 cm};

  \draw[<->] (0,-0.7)--node[below, font=\tiny] {Cross-shore (50 cm)} (5,-0.7);

  \node[font=\small\bfseries] at (2.5,-1.75) {(a) Plan view};

\end{scope}

\begin{scope}[xshift=7.5cm, yshift=0cm]

  \fill[blue!12] (-1.8,-0.9) rectangle (0,2.2);
  \draw[thick, black!45] (-1.8,-0.9)--(5.8,-0.9);
  \node[font=\footnotesize, text=blue!50!black] at (-0.9,0.65) {Water};
  \node[above, font=\tiny, black!55] at (-0.9,-0.9) {Sea bottom};

  \fill[black!45] (0,-0.9) rectangle (5,0);
  \draw[thick, black!55] (0,-0.9) rectangle (5,0);
  \node[font=\tiny, text=white] at (2.5,-0.45) {Solid support};
  \fill[cyan!15] (0,0) rectangle (5,3.0);
  \draw[thick, black!35] (0,0) rectangle (5,3.0);
  \node[font=\footnotesize] at (3.2,2.65) {Ice block};

  \draw[thick, dashed, gray!65] (0,0)--(0,3.0);
  \draw[dashed, blue!55] (-1.8,2.2)--(5.5,2.2);
  \node[right, font=\tiny, blue!55!black] at (5.5,2.2) {MSL ($z{=}0$)};

  \fill[red!75!black] (0.5,1.5) circle (0.13);
  \node[below, font=\tiny] at (0.5,1.35) {T1--T3};
  \draw[gray!60] (0.5,1.63)--(0.5,3.0);
  \fill[red!75!black] (1.7,1.5) circle (0.13);
  \node[below, font=\tiny] at (1.7,1.35) {T4};
  \draw[gray!60] (1.7,1.63)--(1.7,3.0);

  \draw[<->] (4.45,1.5)--(4.45,2.2);
  \node[right, font=\tiny] at (4.48,1.85) {7 cm};
  \draw[<->] (5.05,0)--(5.05,1.5);
  \node[right, font=\tiny] at (5.08,0.75) {$z_b=15$ cm};
  \draw[<->] (5.45,-0.9)--(5.45,0);
  \node[right, font=\tiny] at (5.48,-0.45) {9 cm};
  \draw[<->] (5.8,0)--(5.8,3.0);
  \node[right, font=\tiny] at (5.83,1.5) {30 cm};
  \draw[<->] (-1.55,-0.9)--(-1.55,2.2);
  \node[left, font=\tiny] at (-1.58,0.65) {31 cm};

  \draw[<->] (0,-1.2)--node[below, font=\tiny] {Cross-shore (50 cm)} (5,-1.2);
  \node[font=\small\bfseries] at (2.5,-2.0) {(b) Side view};

\end{scope}

\end{tikzpicture}%
}
\caption{Reconstructed nominal arrangement of the four iButton temperature sensors in
         the 50~cm $\times$ 50~cm $\times$ 30~cm ice block. Coordinates are measured
         from the lower wave-facing corner in the wave-propagation, block-width, and
         vertical directions, respectively. (a)~Plan view: T1 $(5,10,15)$~cm,
         T2 $(5,25,15)$~cm, T3 $(5,40,15)$~cm, and T4 $(17,25,15)$~cm.
         (b)~Side view: all sensors lie 15~cm above the block base. The 9~cm support
         places the sensors 24~cm above the seabed and 7~cm below MSL at the 31~cm
         water depth. Positions are reconstructed from field notes and are approximate.}
\label{fig:ibutton_schematic}
\end{figure}

Four iButton temperature sensors (T1--T4) were embedded in the ice block during mould-filling, with wire leads routed out through the top surface. In block-fixed coordinates measured from the lower wave-facing corner, their nominal positions were T1 $(5,10,15)$~cm, T2 $(5,25,15)$~cm, T3 $(5,40,15)$~cm, and T4 $(17,25,15)$~cm, where the coordinates denote the wave-propagation, block-width, and vertical directions, respectively (Fig.~\ref{fig:ibutton_schematic}). Thus, T1--T3 formed a row 5~cm behind the initial face, while T4 was positioned 12~cm farther into the block behind T2; all sensors were 15~cm above the block base. The 9~cm support base placed the sensors 24~cm above the seabed and therefore 7~cm below MSL in the 31~cm water depth. These positions are reconstructed from field notes and should be regarded as approximate.

\subsubsection{Ice block geometry and properties}
\label{sec:ice_geometry}

Ice blocks were cast from freshwater in rectangular polystyrene moulds with internal dimensions of $0.50$~m (cross-shore depth) $\times$ $0.30$~m (height) $\times$ $0.50$~m (along-shore width). The along-shore width was matched to the internal width of the flume so that the ice block spanned the full cross-section, constraining wave-driven flow to the two-dimensional vertical plane and directing melt recession in the cross-shore direction only. The iButton sensor wire leads were cast in place during the freezing process
(Fig.~\ref{fig:procedure}a).

The ice blocks and flume water were freshwater. The model calculations retained the reference thermophysical constants listed in Table~\ref{tab:ice_properties}; these values are part of the reconstructed model specification and were not direct measurements of the flume water. Their effect is therefore included in the residual uncertainty absorbed by the calibration.

\begin{table}[htbp]
    \centering
    \caption{Reference physical properties used in the melt-rate formulation.}
    \label{tab:ice_properties}
    \begin{tabular}{llcl}
        \hline
        Symbol & Quantity & Value & Unit \\
        \hline
        $\rho_i$       & Ice density                       & $917$                & kg\,m$^{-3}$ \\
        $\Gamma$        & Latent heat of fusion             & $3.34 \times 10^{5}$ & J\,kg$^{-1}$ \\
        $\rho_w$        & Reference water density           & $1025$               & kg\,m$^{-3}$ \\
        $c_p$           & Reference water heat capacity     & $3985$               & J\,kg$^{-1}$\,K$^{-1}$ \\
        $\mathrm{Pr}$   & Prandtl number                    & $13$                 & --- \\
        $\nu$           & Kinematic viscosity of water      & $1.30 \times 10^{-6}$ & m$^{2}$\,s$^{-1}$ \\
        $k_s$           & Ice-surface roughness length      & $0.01$               & m$^{\dagger}$ \\
        \hline
        \multicolumn{4}{p{0.88\textwidth}}{\small $^{\dagger}$Adopted from \citet{white1980} for consistency; not measured in the present tests.}
    \end{tabular}
\end{table}

\subsubsection{Wave conditions}
\label{sec:wave_conditions}

Regular monochromatic waves were generated for both test cases, targeting a wave height of $H_\mathrm{target} = 0.05$~m in each case. The two cases were designed to span qualitatively different wave-forcing regimes primarily through their periods.

Test~1 employed longer-period waves ($T = 1.54$~s, $f = 0.65$~Hz). Wave run-up at the ice face was gentle and orbital velocities were moderate, resulting in comparatively calm melting conditions with limited turbulent mixing.

Test~2 used shorter-period waves ($T = 0.87$~s, $f = 1.15$~Hz). The shorter wavelength produced a more energetic wave field at the ice front, characterised by vigorous run-up and stronger near-surface orbital velocities. Intermittent local breaking was observed at the ice face. Both tests were run for a duration of $1800$~s at a constant water depth of $d = 0.31$~m. The measured wave conditions, including effective heights and reflection coefficients, are reported in Section~\ref{sec:measured_waves}.

\subsubsection{Test matrix}
\label{sec:test_matrix}

A total of four runs were attempted during the experimental campaign. Table~\ref{tab:test_matrix} summarises all runs, including their wave parameters and status. Tests~0 and~3 are excluded from the quantitative analysis: Test~0 served as a pilot run to verify instrumentation and data acquisition procedures, while Test~3 was designed to explore irregular wave conditions but was terminated due to equipment failure before usable data could be obtained.

\begin{table}[htbp]
    \centering
    \caption{Test matrix for the wave-induced melting experiment. $T$: target wave period;
             $H_\mathrm{target}$: target generator wave height; $d$: water depth.
             Measured effective wave heights and temperature differences are reported
             in Section~\ref{sec:results}.}
    \label{tab:test_matrix}
    \begin{tabular}{ccccc}
        \hline
        Test & $T$ (s) & $H_\mathrm{target}$ (m) & $d$ (m) & Status \\
        \hline
        0 & ---  & ---   & 0.31 & Pilot run (no analysis) \\
        1 & 1.54 & 0.050 & 0.31 & Valid \\
        2 & 0.87 & 0.050 & 0.31 & Valid \\
        3 & ---  & ---   & 0.31 & Failed (data loss) \\
        \hline
    \end{tabular}
\end{table}

\subsubsection{Data processing}
\label{sec:data_processing}

\paragraph{Ice-front profile extraction.}
The position of the ice front was extracted from the recorded video using a five-step image processing pipeline implemented in MATLAB. In Step~1, each video frame was geometrically rectified using a projective transformation estimated from the four corner points of the orange flume frame visible in the image. The transformation was computed using \texttt{fitgeotrans} with a projective model and applied via \texttt{imwarp} to produce an undistorted, rectified view of the ice face. In Step~2, we manually digitised the visible ice--water interface by selecting points along the ice front. In Step~3, the physical scale of the rectified image was established from the two rulers visible in the camera view: the white horizontal ruler provided the horizontal scale factor $s_x$, and the yellow vertical ruler provided the vertical scale factor $s_y$ (cm\,px$^{-1}$). In Step~4, we identified the pixel column corresponding to $x = 0$ (the initial ice face position) using a fixed background reference marker, and the pixel row corresponding to $z = 0$ (mean still-water level). In Step~5, these reference coordinates were used to transform all digitised ice-front pixels from image space into the physical coordinate system $(X,\,Z)$, with $X$ positive in the cross-shore direction away from the initial face and $Z$ positive upward from mean still-water level, according to
\begin{equation}
    X = (p_x - x_0)\,s_x, \qquad Z = (y_0 - p_y)\,s_y,
    \label{eq:coord_transform}
\end{equation}
where $(p_x,\,p_y)$ are the pixel coordinates of a digitised point, $x_0$ is the pixel column corresponding to $X = 0$, and $y_0$ is the pixel row corresponding to $Z = 0$.

\begin{figure}[htbp]
    \centering
    \includegraphics[width=\textwidth]{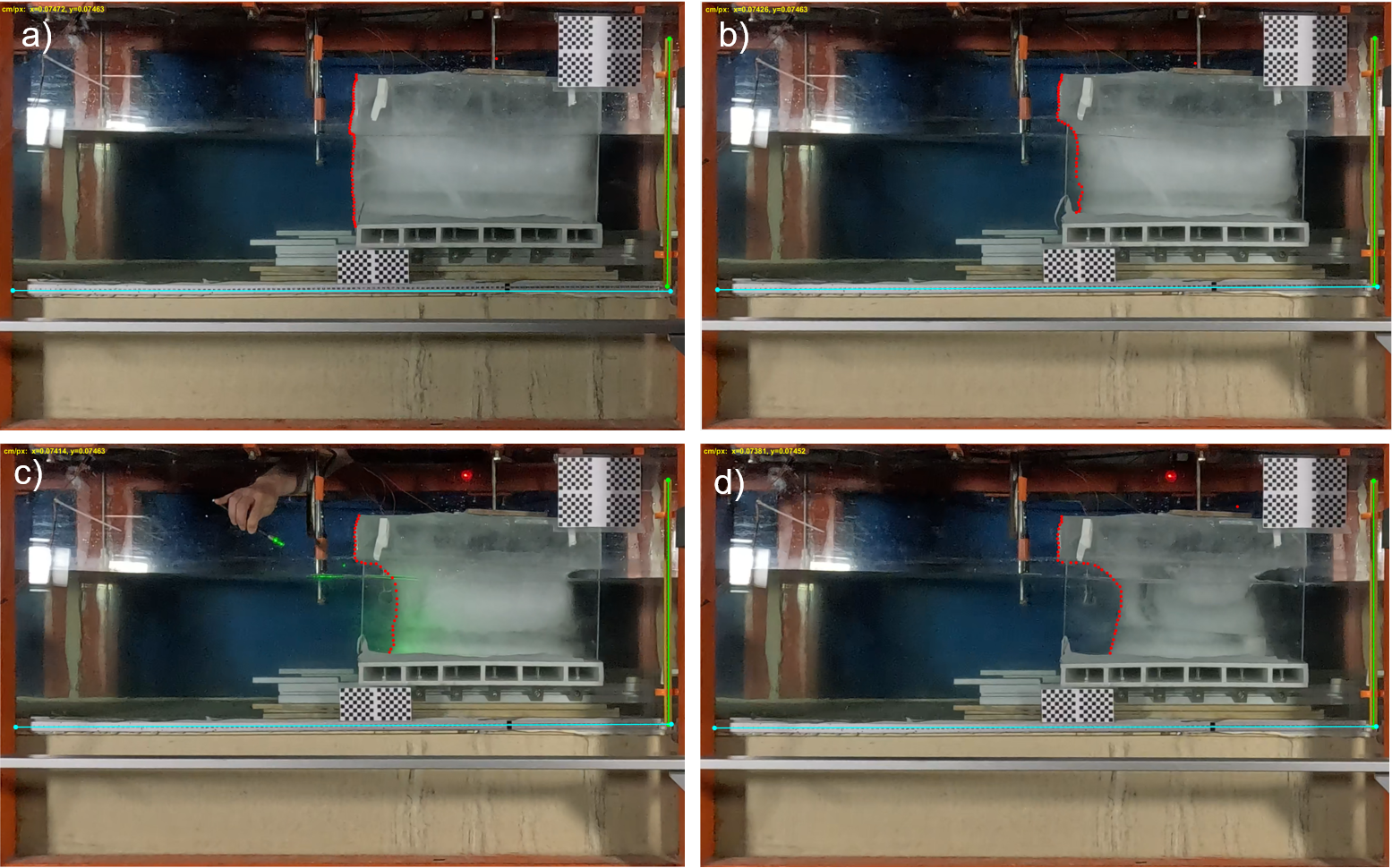}
    \caption{Manually digitised ice--water interface for Test~1 at (a)~$t=0$, (b)~$t=10$~min,
             (c)~$t=20$~min, and (d)~$t=30$~min. Red dots mark selected interface points.
             The horizontal blue line (lower edge) and vertical green line (left edge) are the
             physical reference lengths used to compute the horizontal and vertical scale
             factors $s_x$ and $s_y$ (cm\,px$^{-1}$); the corresponding values are printed in
             the upper-left corner of each panel.}
    \label{fig:interface_tracking}
\end{figure}

Figure~\ref{fig:interface_tracking} shows representative rectified frames from Test~1 at $t = 0$, 10, 20, and 30~min, illustrating the manual interface digitisation. Red dots mark the selected ice--water interface points; the horizontal blue line and vertical green line indicate the physical reference lengths used to establish the scale factors $s_x$ and $s_y$ respectively, with the corresponding numerical values printed in the upper-left corner of each panel. The manual cursor-placement error in step~2 is estimated at approximately 3--5 pixels per point. For the Test~1 frames shown here, the stored MATLAB calibration gives $s_x = 0.738$--$0.747$~mm\,px$^{-1}$ and $s_y = 0.745$--$0.746$~mm\,px$^{-1}$, so this cursor-placement uncertainty corresponds to approximately 2.2--3.7~mm. This is well below the smallest measured recession increment and does not materially affect the profile comparisons.

\section{Results}
\label{sec:results}

\subsection{Experimental observations}
\label{sec:experimental_results}

\subsubsection{Measured wave conditions}
\label{sec:measured_waves}

Wave conditions were characterised by applying the three-gauge incident--reflected decomposition of \citet{mansardfunke1980} to each gauge group separately. Table~\ref{tab:wave_conditions} reports the results for the gauge group closest to the ice block (Group~2), which provides the closest available estimate of the local wave field near the ice face.

\begin{table}[htbp]
    \centering
    \caption{Measured wave conditions from the near-ice gauge group (Group~2) using the
             \citet{mansardfunke1980} decomposition. Target generator
             setting was $H = 0.05$~m for both tests. $H_i$: incident wave height;
             $H_r$: reflected wave height; $R$: reflection coefficient;
             $H_\mathrm{eff} = H_i + H_r$: effective wave-height proxy used at the ice face.}
    \label{tab:wave_conditions}
    \begin{tabular}{ccccc}
        \hline
        Test & $H_i$ (m) & $H_r$ (m) & $R$ (\%) & $H_\mathrm{eff}$ (m) \\
        \hline
        1 ($T = 1.54$~s) & 0.044 & 0.036 & 83 & 0.080 \\
        2 ($T = 0.87$~s) & 0.068 & 0.049 & 72 & 0.117 \\
        \hline
    \end{tabular}
\end{table}

For Test~1, the incident wave height near the ice ($H_i = 0.044$~m) was close to the target generator setting of 0.05~m. For Test~2, the measured near-ice incident height ($H_i = 0.068$~m) exceeded the target by 36\%. Consequently, the measured rather than target wave conditions are used below.

The local wave elevation near the reflecting ice face contains both incident and reflected components. We therefore use $H_\mathrm{eff}=H_i+H_r$ as the wave-height proxy for calculating the orbital kinematics at the interface. This gives $H_\mathrm{eff}=0.080$~m for Test~1 and $0.117$~m for Test~2. Because Group~2 was not co-located with the ice--water interface, these values do not resolve the detailed standing-wave field at the face; the resulting uncertainty is discussed in Section~\ref{sec:limitations}.

In both tests the waves were confirmed to be free of depth-limited breaking at the wave gauges, consistent with the incident-wave assumption underlying the White~(1980) framework (Section~\ref{sec:white_framework}). Test~1 also showed no visible breaking at the ice face. In Test~2, however, wave crests intermittently impacted the ice face; this was a local ice-face interaction rather than depth-induced shoaling breaking, and is discussed in Section~\ref{sec:limitations} (L2).

\subsubsection{Visual comparison of melting behaviour}
\label{sec:visual_comparison}

\begin{figure}[htbp]
    \centering
    \includegraphics[width=\textwidth]{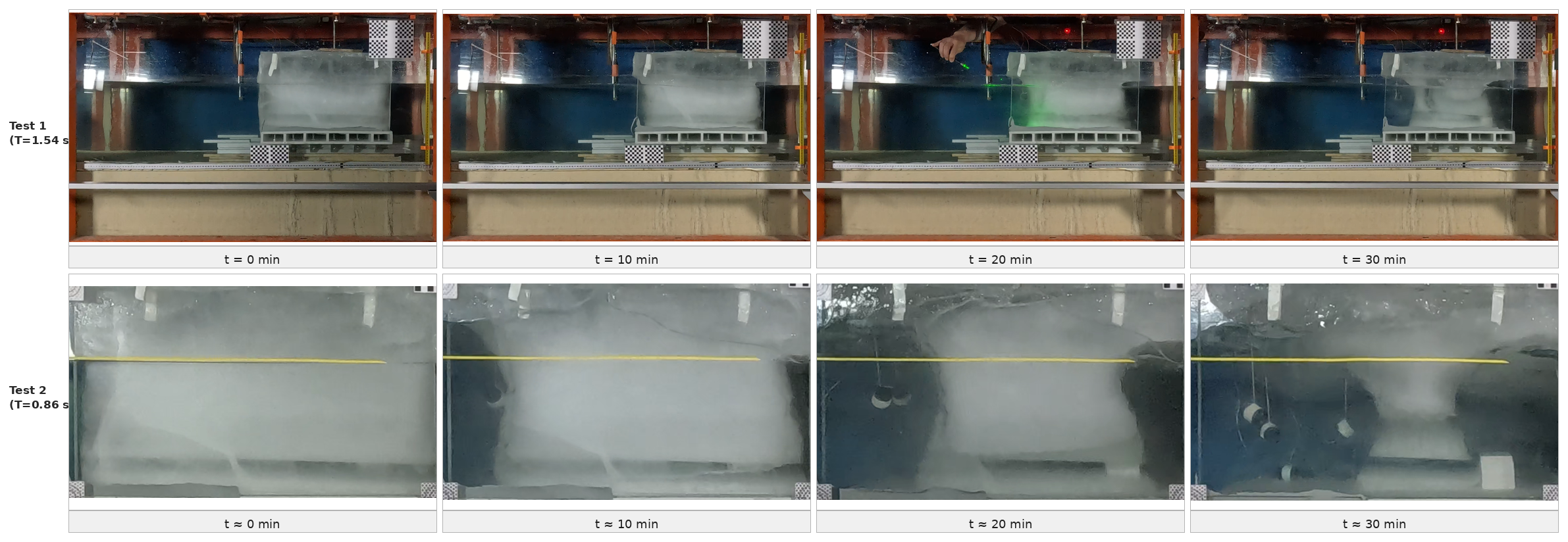}
    \caption{Ice-front evolution over 30~minutes. Top row: Test~1 ($T=1.54$~s, long waves);
             bottom row: Test~2 ($T=0.87$~s, short waves). Each column corresponds to
             $t=0$, 10, 20 and 30~min. In Test~1 the ice face retreats gradually and
             near-uniformly with depth, whereas in Test~2 stronger wave run-up and
             intermittent ice-face impact accelerate melting.}
    \label{fig:melting_progression}
\end{figure}

Figure~\ref{fig:melting_progression} shows snapshots of the ice face at $t=0$, 10, 20 and 30~min for both tests, arranged for direct visual comparison.

Under Test~1 conditions, wave run-up at the ice face was moderate and the ice retreated in a relatively smooth, depth-graded manner. The free surface remained calm between wave cycles, and recession was gradual along the visible ice--water interface. By contrast, Test~2 produced more energetic wave action at the ice face. Higher orbital velocities, shorter wave period, and stronger run-up generated more vigorous near-surface mixing and intermittent ice-face impact. Under these conditions, ice loss was substantially faster. This difference is consistent with the expected scaling of the wave-induced melt rate with the velocity scale $H/T$ and with enhanced convective renewal of water near the ice face.

\subsubsection{Temperature evolution}
\label{sec:temperature_results}

\begin{figure}[htbp]
    \centering
    \begin{minipage}[t]{0.49\textwidth}
        \includegraphics[width=\textwidth]{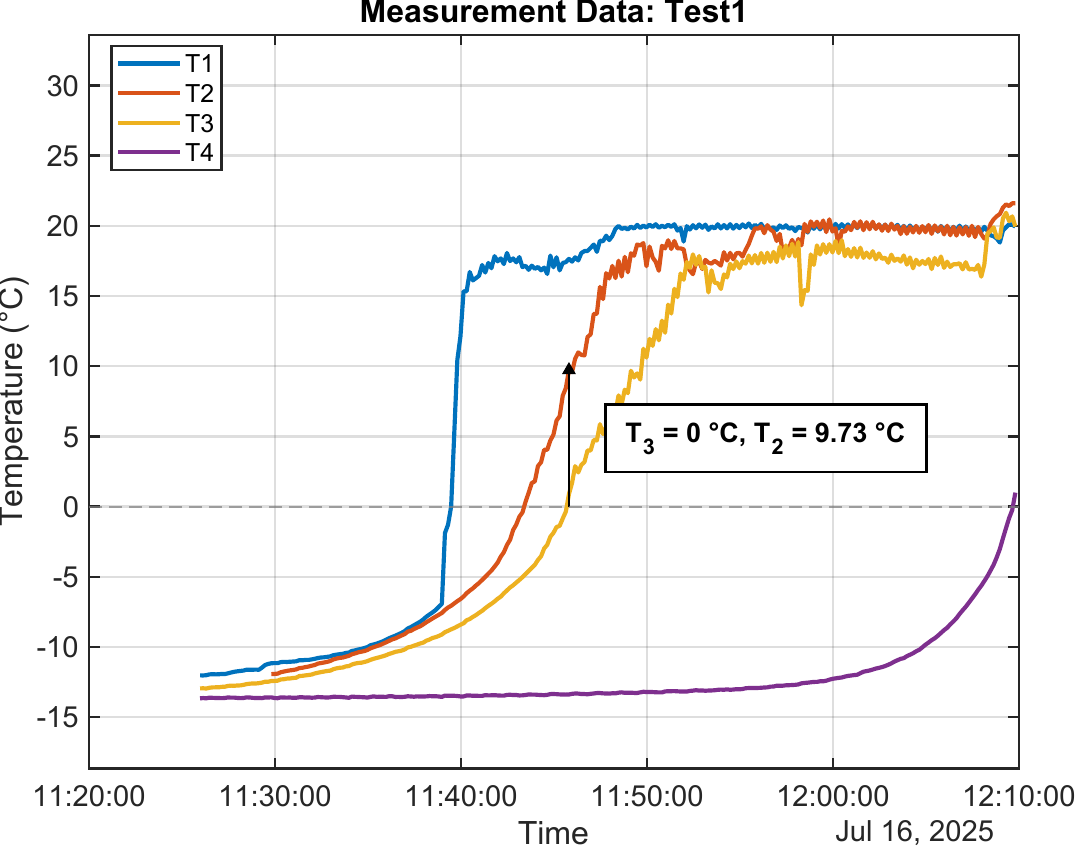}
    \end{minipage}
    \hfill
    \begin{minipage}[t]{0.49\textwidth}
        \includegraphics[width=\textwidth]{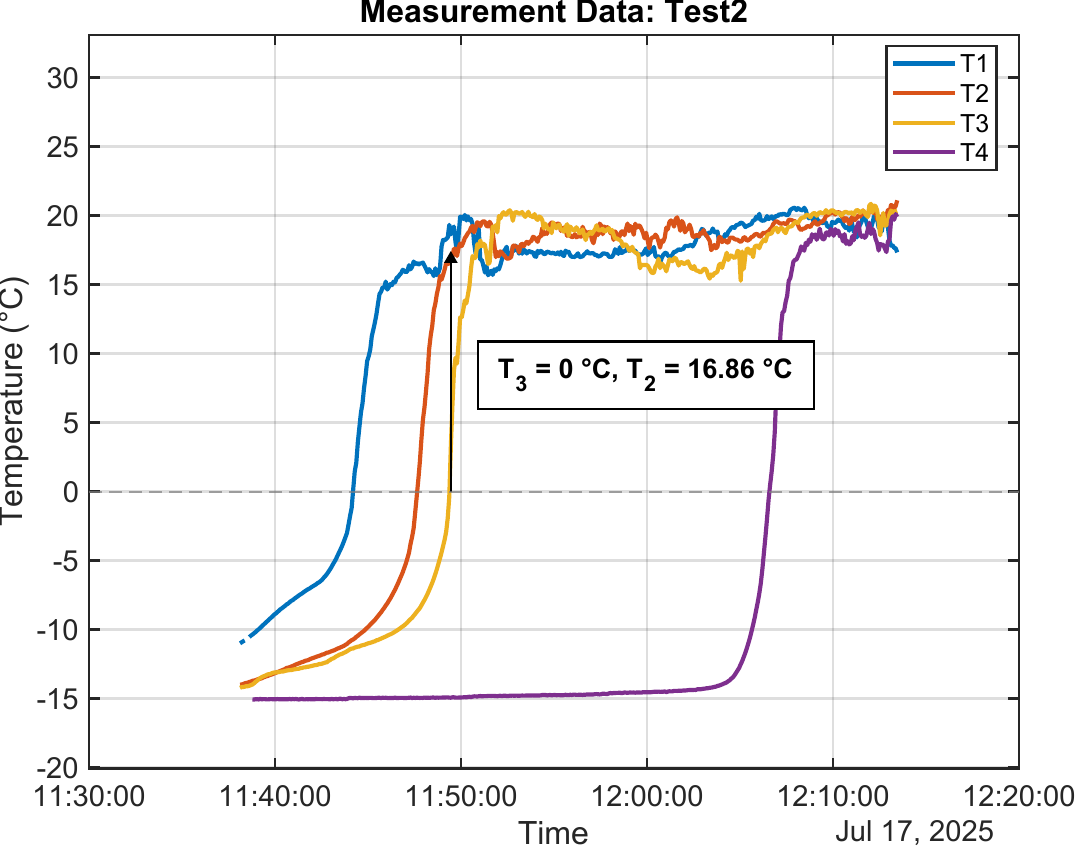}
    \end{minipage}
    \caption{iButton temperature records for Test~1 (left, $T=1.54$~s) and Test~2 (right,
             $T=0.87$~s). The vertical arrow marks the moment $t^*$ at which T3
             first reaches $0\,^\circ$C; the annotated value gives the concurrent T2
             reading used as the proxy for the near-ice temperature difference $\Delta T$:
             $9.73\,^\circ$C in Test~1 and $16.86\,^\circ$C in Test~2.}
    \label{fig:temperature_evolution}
\end{figure}

Figure~\ref{fig:temperature_evolution} shows the iButton temperature records for both tests. As the ice face retreated toward the sensor array, the near-front row T1--T3, which shared the same cross-shore and vertical coordinates but differed in along-shore position, became exposed at slightly different times because the melt front was not perfectly uniform across the block width. T4 was positioned 12~cm farther into the block. The instant at which T3 first crosses $0\,^\circ$C marks a reference moment $t^*$ used to estimate $\Delta T$ (see below).

The temperature records differ substantially between the two tests. When T3 first reached $0\,^\circ$C, T2 read $9.73\,^\circ$C in Test~1 and $16.86\,^\circ$C in Test~2. The records are consistent with different exposure and mixing histories, but the absence of a co-located ambient-water record prevents the difference from being attributed uniquely to wave-driven mixing.

In the absence of a dedicated ambient-water temperature sensor, the near-ice temperature difference $\Delta T$ required by the White~(1980) formulation was derived directly from the iButton records. The instant at which sensor $T_3$ first registers $0\,^\circ$C was identified as a reference moment $t^*$ at which this sensor transitions from frozen-in to water-exposed. At this moment, the adjacent sensor $T_2$ had already been exposed and its reading reflects the near-surface water temperature. Since $T_2$ and $T_3$ were placed at the same cross-shore and vertical position but at different along-shore locations, this timing offset reflects minor three-dimensional variation in the melt front; $\Delta T \approx T_2(t^*)$ is adopted as the best available single-point proxy for the near-ice thermal forcing. The value was therefore taken as a single representative estimate for each test, yielding
\begin{equation*}
\Delta T = 9.73\,^\circ\text{C} \quad (\text{Test~1}), \qquad \Delta T = 16.86\,^\circ\text{C} \quad (\text{Test~2}).
\end{equation*}
The physical significance and limitations of this approximation---in particular the dependence of the proxy on sensor exposure and wave conditions rather than on independently controlled boundary conditions---are discussed further in Section~\ref{sec:temperature_importance}.

\subsection{Results comparison and velocity parameterization}
\label{sec:results_comparison}


\subsubsection{Theoretical predictions for $\alpha = 1$}
\label{sec:alpha_benchmark}

\begin{figure}[htbp]
    \centering
    \includegraphics[width=0.85\textwidth,trim=0 0 0 22bp,clip]{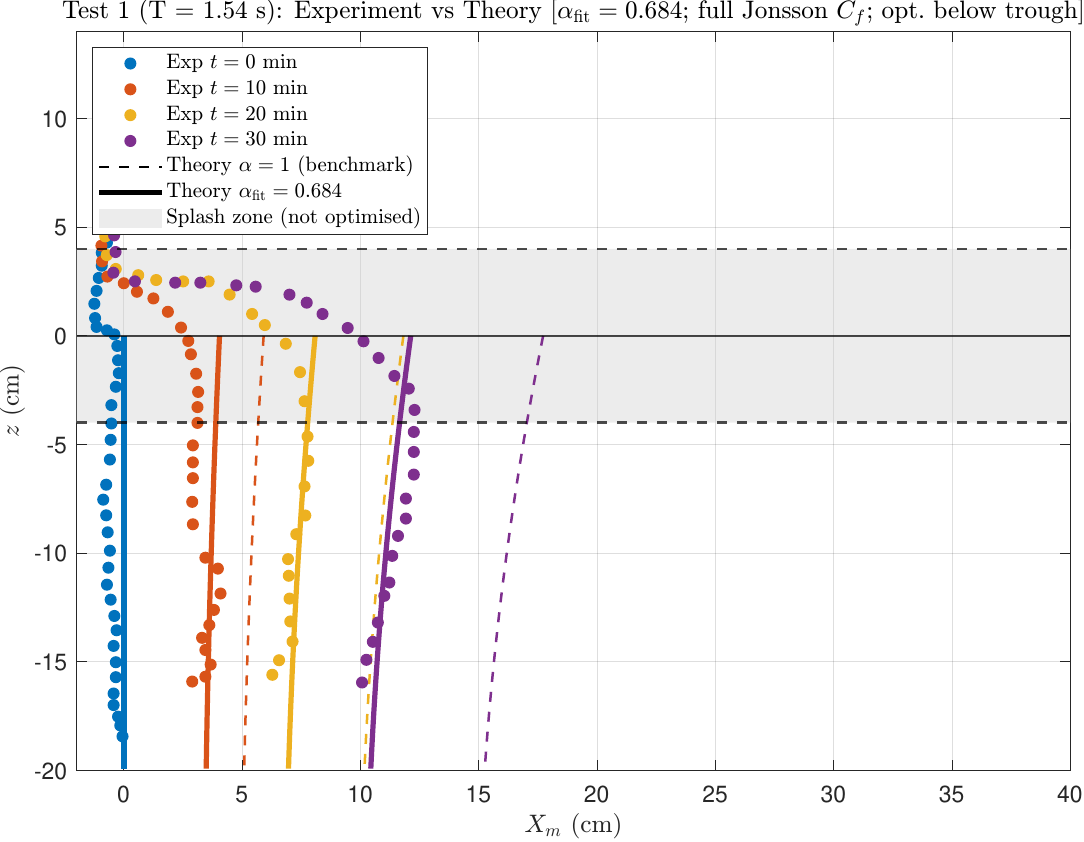}
    \caption{Test~1 ($T = 1.54$~s): measured ice-front profiles (dots) and theoretical
             erosion $X_m(z,t)$ from the modified formulation.
             Dashed lines: $\alpha = 1$ (benchmark);
             solid lines: calibrated $\alpha_\mathrm{fit} = 0.684$.
             Grey band: splash zone (wave trough to crest); dashed horizontal lines
             mark the trough ($z = -H/2$) and crest ($z = +H/2$) depths.}
    \label{fig:comparison_test1}
\end{figure}

\begin{figure}[htbp]
    \centering
    \includegraphics[width=0.85\textwidth,trim=0 0 0 22bp,clip]{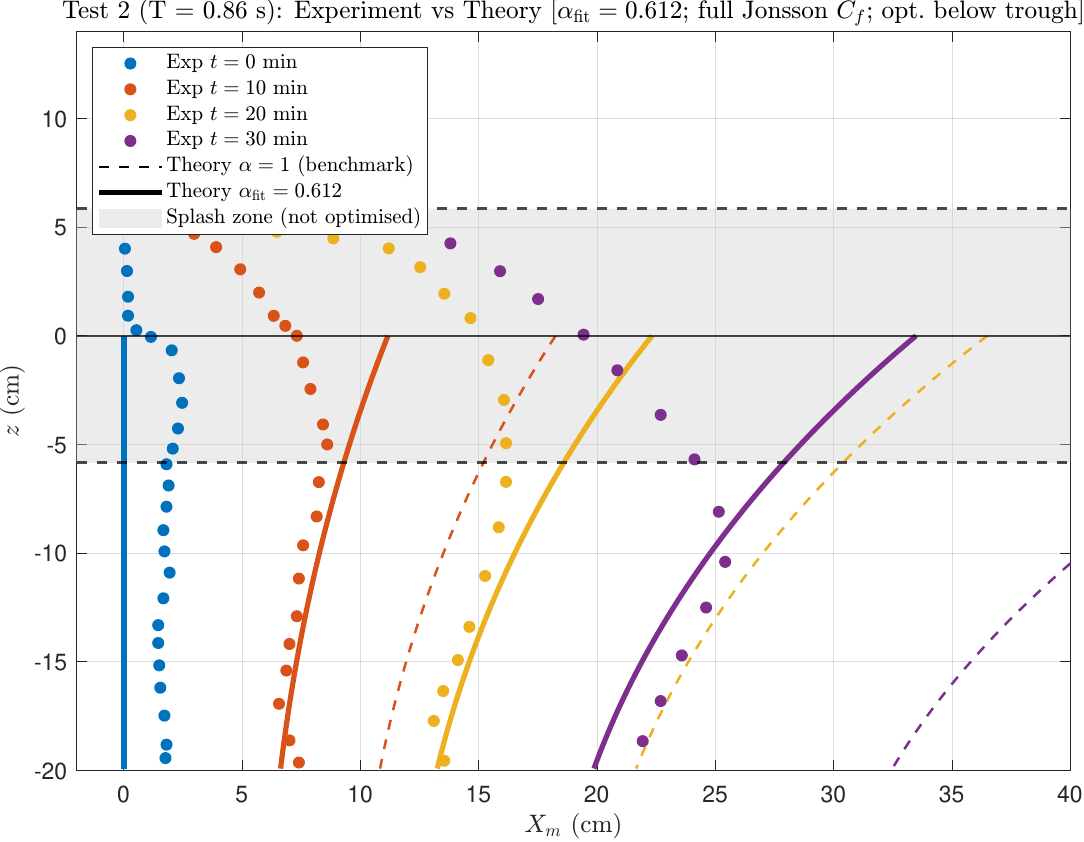}
    \caption{As Fig.~\ref{fig:comparison_test1}, but for Test~2 ($T = 0.87$~s);
             calibrated $\alpha_\mathrm{fit} = 0.612$.}
    \label{fig:comparison_test2}
\end{figure}

Figures~\ref{fig:comparison_test1} and~\ref{fig:comparison_test2} show the measured ice-front profiles (dots) overlaid on the theoretical erosion profiles at $t = 0$, 10, 20, and 30~min. In each figure the grey-shaded band marks the splash zone (between the wave trough and crest), where intermittent submergence and wave run-up complicate the direct interpretation of the boundary-layer formulation; dashed lines show the $\alpha = 1$ benchmark and solid lines the calibrated $\alpha_\mathrm{fit}$ discussed below.

At $\alpha = 1$ the theoretical profiles systematically overpredict the observed recession below mean sea level for both tests. The present data cannot separate contributions from the velocity closure, thermal forcing, assumed roughness and water properties, or the wave-height proxy; their combined effect is represented empirically by $\alpha$.

\subsubsection{Calibration of $\alpha$ by least-squares fitting}
\label{sec:alpha_calibration}

The coefficient $\alpha$ is determined for each test by minimising the sum of squared differences between the measured and predicted erosion profiles. To avoid contaminating the calibration with the splash-zone physics discussed in Section~\ref{sec:splash_zone}, only experimental data points below the wave trough ($z < -H/2$) are included:
\begin{equation}
\alpha_\mathrm{fit}
= \arg\min_{\alpha}\,
\sum_{j}\sum_{i:\,z_i < -H/2}
\left[X_m^\mathrm{exp}(z_i,\,t_j) - X_m^\mathrm{th}(z_i,\,t_j;\,\alpha)\right]^2 .
\label{eq:lsq_alpha}
\end{equation}
The minimisation uses \texttt{fminsearch} in MATLAB. The resulting best-fit values are
\begin{equation}
\alpha_\mathrm{fit} = 0.684 \quad (\text{Test~1}), \qquad
\alpha_\mathrm{fit} = 0.612 \quad (\text{Test~2}).
\label{eq:alpha_results}
\end{equation}

Both values are below unity, consistent with the overprediction at $\alpha = 1$ described above. After accounting for the measured wave height, period, and temperature forcing in the model, the two fitted values remain similar, suggesting that $\alpha$ captures a systematic residual closure over the tested range. A broader dataset spanning wider ranges of $H$, $T$, and $\Delta T$ would be needed to establish regime-independence rigorously. The physical interpretation of $\alpha < 1$ and its implications for field-scale application are discussed in Section~\ref{sec:turbulence_importance}.

\section{Discussion}
\label{sec:discussion}

\subsection{Role of temperature forcing}
\label{sec:temperature_importance}

The driving temperature difference $\Delta T$ enters Eq.~\eqref{eq:Vm_full} as a linear factor and is therefore the single most influential scalar parameter in the formulation. In the present experiments no dedicated ambient water temperature sensor was deployed; instead $\Delta T$ was approximated as the temperature recorded by T2 at the moment T3 first crossed 0\textdegree C, as described in Section~\ref{sec:temperature_results}. This proxy captures a snapshot of the near-ice water temperature at the most active phase of melting, but it cannot resolve the full time evolution of the thermal boundary layer during the experiment.

A particularly important consequence is that $\Delta T$ was not an independent experimental control variable. The adopted proxies are $\Delta T=9.73\,^\circ$C for Test~1 and $16.86\,^\circ$C for Test~2 (Section~\ref{sec:temperature_results}). They may reflect differences in wave-driven mixing, sensor exposure history, and ambient conditions. These effects cannot be separated with the present dataset, and the reference water properties used in the model were not measured during the tests.

This coupling between the wave-induced velocity field and $\Delta T$ represents an important limitation for the present dataset. The calibration parameter $\alpha$ may therefore absorb some of the uncertainty in the temperature forcing in addition to the residual velocity-scale correction discussed below. Resolving this ambiguity requires simultaneous, spatially resolved temperature measurements in the near-ice boundary layer---a key motivation for the improved instrumentation proposed for future work.

\subsection{Role of turbulence and implications for the velocity scaling}
\label{sec:turbulence_importance}

Both calibrated values ($\alpha_\mathrm{fit} = 0.684$, $0.612$) are below unity, consistent with the systematic overprediction at $\alpha = 1$ identified in Section~\ref{sec:alpha_benchmark}. The parameter $\alpha$ therefore quantifies the residual velocity-scale and experimental-forcing uncertainty within the modified formulation. The fitted values differ by approximately 11\% relative to their mean. Their similarity is encouraging, but two tests are insufficient to establish that the residual correction is independent of wave regime.

The product $\alpha_\mathrm{fit} C'_{\rm ours} \approx 1.28$--$1.43\times10^{-4}$ falls close to White's published coefficient of $1.46\times10^{-4}$. This comparison shows that the fitted effective magnitude is similar to White's value, but it does not resolve the undocumented step in White's derivation or provide an independent validation because $\alpha$ was fitted to the present data. The reconstruction instead distinguishes the documented closure chain from the empirical adjustment required for these experiments.

The incident wave field in both experiments was within the non-breaking regime at the wave gauges. After calibration, the formulation represents the observed below-trough depth dependence for the two tested conditions (Figs.~\ref{fig:comparison_test1} and~\ref{fig:comparison_test2}). Breaking-wave turbulence, local impact, ice-face roughness evolution, and subsurface buoyancy-driven circulation remain outside the formulation and require dedicated future study.

\subsection{Limitations of the current study}
\label{sec:limitations}

The principal assumptions and experimental constraints are summarised in Table~\ref{tab:limitations}. The dominant limitations are the linear, non-breaking wave framework and the lack of independently controlled, spatially resolved temperature measurements near the ice face.

\begin{table}[htbp]
\centering
\caption{Key limitations, their effect on the results, and recommended mitigations for future work.}
\label{tab:limitations}
\small
\begin{tabular}{>{\raggedright\arraybackslash}p{0.03\linewidth}
                >{\raggedright\arraybackslash}p{0.25\linewidth}
                >{\raggedright\arraybackslash}p{0.32\linewidth}
                >{\raggedright\arraybackslash}p{0.25\linewidth}}
\hline
 & Limitation & Effect on results & Future mitigation \\
\hline
L1 & Linear wave theory & Inapplicable to breaking, nonlinear, or irregular waves &
     Nonlinear, breaking-wave, and spectral extension \\
L2 & Breaking and splash-zone turbulence not resolved &
     Boundary-layer heat-transfer model cannot represent impact, aeration, and intermittent contact above the trough &
     Dedicated breaking/run-up measurements and turbulence-resolving parameterization \\
L3 & $\Delta T$ proxy, approximate iButton positions, and reference water properties &
     Thermal-forcing uncertainty; $\alpha$ may absorb temperature and property errors &
     Thermistor chain at known positions; measured ambient temperature and properties \\
L4 & Wave gauges not co-located with ice face &
     $H_\mathrm{eff}$ uncertain at ice--water interface &
     Co-located gauge or video-based elevation extraction \\
L5 & Limited experimental matrix &
     Period, amplitude, roughness, and temperature effects cannot yet be separated fully &
     Additional controlled runs spanning one parameter at a time \\
\hline
\end{tabular}
\end{table}

Regarding L2, a physically distinct effect was observed in Test~2: as the ice face evolved under sustained wave attack, wave crests intermittently impacted the upper part of the face directly, producing visible aeration and turbulent mixing. This is not depth-induced shoaling breaking (the flume bed is flat) but a localised ice-face impact process. It is therefore outside the smooth, continuously submerged boundary-layer idealisation and may contribute to the larger residuals in Test~2 relative to Test~1.

Future campaigns should address thermal instrumentation as a priority. A dedicated ambient-water temperature sensor logging continuously throughout the test would provide a stable reference for the bulk thermal forcing. A multi-point thermistor chain deployed at known, fixed distances from the initial ice face (e.g.\ at 1, 3, 5, and 10~cm) would resolve the temporal evolution of the near-ice thermal field and allow $\Delta T$ to be estimated directly as the face retreats. Controlled initial water temperature and active monitoring would not eliminate the near-ice thermal boundary layer, but would make the thermal forcing sufficiently constrained to separate its influence from the velocity scaling parameter $\alpha$ more clearly.

\subsection{Breakdown of the White formulation in the splash zone}
\label{sec:splash_zone}

\begin{figure}[htbp]
    \centering
    \includegraphics[width=0.85\textwidth,trim=0 0 0 22bp,clip]{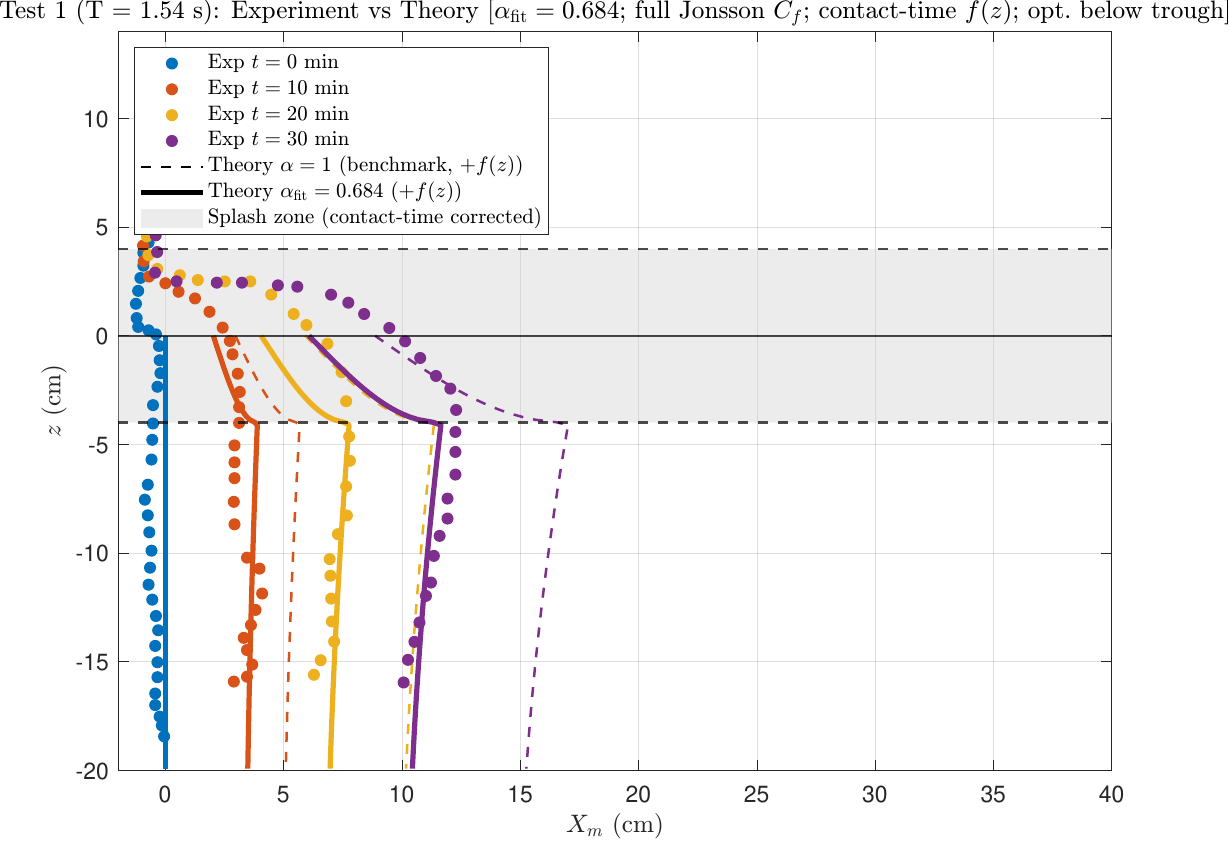}
    \caption{Test~1: as Fig.~\ref{fig:comparison_test1}, but with the contact-fraction
             diagnostic $f(z)$ applied throughout the splash zone
             (Eq.~\eqref{eq:contact_fraction}).}
    \label{fig:comparison_test1_disc}
\end{figure}

\begin{figure}[htbp]
    \centering
    \includegraphics[width=0.85\textwidth,trim=0 0 0 22bp,clip]{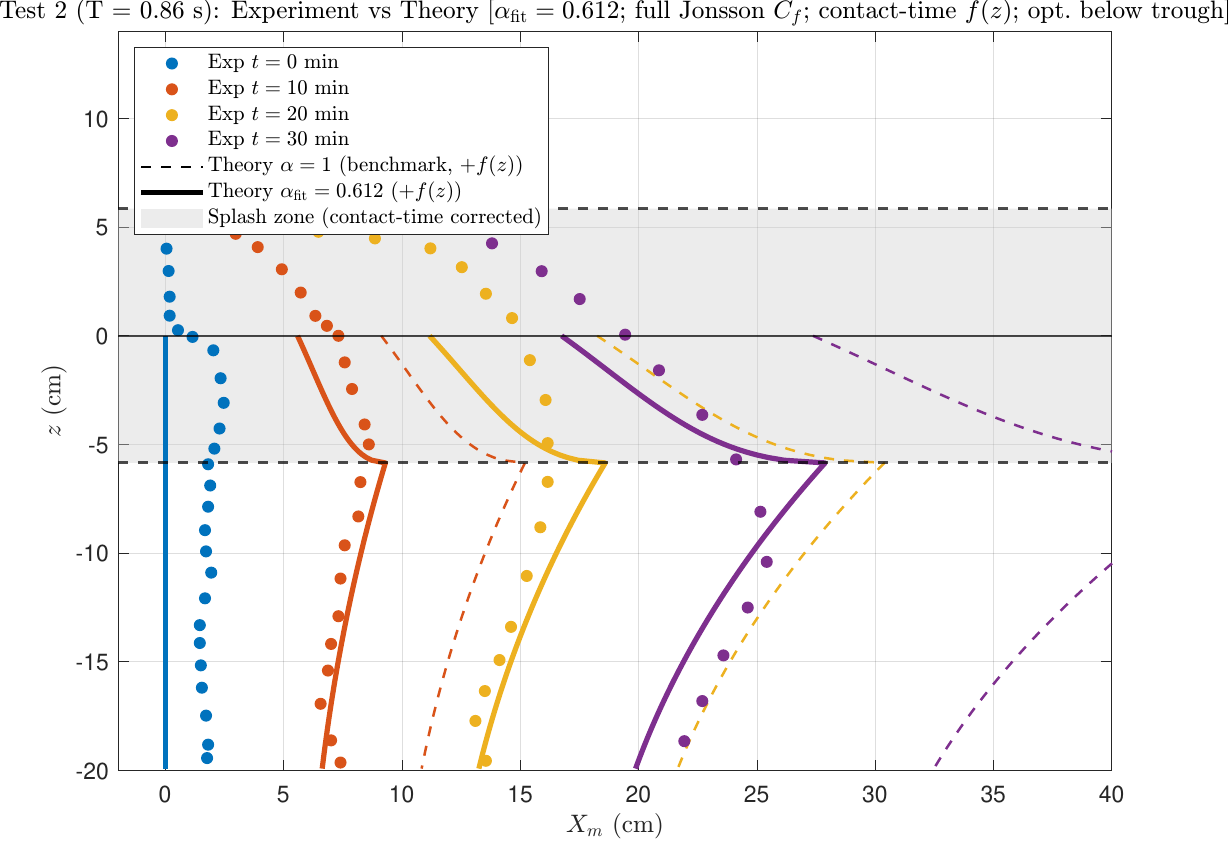}
    \caption{As Fig.~\ref{fig:comparison_test1_disc}, but for Test~2.}
    \label{fig:comparison_test2_disc}
\end{figure}

The measured ice-front profiles show that the maximum horizontal recession occurs not at the mean water level ($z = 0$), as the unmodified White~(1980) formulation predicts, but at a depth several centimetres below it---broadly consistent with the wave trough. This behaviour is expected because the ice face is only intermittently submerged in the splash zone between wave trough ($z = -H/2$) and wave crest ($z = +H/2$). As a simple diagnostic for the present melt profiles, the submerged fraction of a point at elevation $z$ under a sinusoidal surface elevation is
\begin{equation}
f(z) = \frac{1}{\pi}\arccos\!\left(\frac{2z}{H}\right), \qquad -\frac{H}{2} \leq z \leq \frac{H}{2}.
\label{eq:contact_fraction}
\end{equation}
At the mean water level $f(0) = 1/2$; at the trough $f(-H/2) = 1$; and at the crest $f(+H/2) = 0$. Multiplying Eq.~\eqref{eq:Vm_full} by this contact fraction throughout the splash zone is not introduced here as a new melt model, but as a compact way to interpret why the observed profile rolls off toward the waterline. The resulting profiles are shown in Figs.~\ref{fig:comparison_test1_disc} and~\ref{fig:comparison_test2_disc}.

Figures~\ref{fig:comparison_test1_disc} and~\ref{fig:comparison_test2_disc} show the theoretical profiles after applying the diagnostic factor $f(z)$. The solid lines roll off smoothly toward zero as $z \to +H/2$, qualitatively reproducing the experimentally observed peak-below-MSL shape. Table~\ref{tab:waterline_diff} quantifies the remaining discrepancy at the mean water level ($z = 0$) after this diagnostic adjustment is applied.

\begin{table}[htbp]
\centering
\caption{Absolute difference between the contact-fraction-adjusted theoretical prediction
         and the interpolated experimental measurement at the mean water level ($z = 0$),
         for each time snapshot. Theory values from Eq.~\eqref{eq:Vm_full} with
         diagnostic factor $f(z)$ (Eq.~\eqref{eq:contact_fraction}) and $\alpha_\mathrm{fit}$
         from Eq.~\eqref{eq:alpha_results}.}
\label{tab:waterline_diff}
\resizebox{\textwidth}{!}{%
\begin{tabular}{lcccc}
\toprule
Test & $t$ (min) & $X_m^{\rm th}(0,t)$ (cm) & $X_m^{\rm exp}(0,t)$ (cm) & $|\Delta X_m|$ (cm) \\
\midrule
Test~1 ($T = 1.54$~s) & 10 & 2.02 & 2.60 & 0.58 \\
                       & 20 & 4.04 & 6.47 & 2.42 \\
                       & 30 & 6.06 & 9.85 & 3.79 \\
\midrule
Test~2 ($T = 0.87$~s) & 10 &  5.58 &  7.31 & 1.73 \\
                       & 20 & 11.16 & 14.96 & 3.80 \\
                       & 30 & 16.74 & 19.47 & 2.73 \\
\bottomrule
\end{tabular}
}
\end{table}

The absolute differences are modest in absolute terms: for Test~2 at $t = 30$~min the theory falls within 2.73~cm of the experiment at the waterline, where the total measured recession is 19.47~cm---a relative error of 14\%. For Test~1 the percentage error is larger (38\% at $t = 30$~min) but the absolute discrepancy of 3.79~cm is comparable in magnitude to Test~2 and must be interpreted against the smaller total erosion of 9.85~cm. Notably, for Test~2 the discrepancy does not grow monotonically with time: it peaks at $t = 20$~min (3.80~cm) before reducing at $t = 30$~min (2.73~cm), a behaviour that may reflect the evolving thermal boundary layer discussed in Section~\ref{sec:temperature_importance}.

The contact-fraction factor $f(z)$ is only a diagnostic adjustment. It accounts for intermittent submersion but not the additional run-up, aeration, impact, and turbulent transport processes present in the splash zone. Those processes remain outside the White (1980) boundary-layer framework and require dedicated measurements and modelling.

\section{Conclusions}
\label{sec:conclusions}

This study reconstructs White's (1980) wave-induced ice-melt formulation and extends its rough-wall closure over depth. The resulting velocity coefficient is calibrated against controlled laboratory experiments. The principal findings are as follows.

\textbf{(1) Incomplete derivation chain and coefficient discrepancy.} The published waterline coefficient $1.46 \times 10^{-4}$ cannot be recovered from the documented closure chain alone. White's resultant-velocity definition gives $C' \approx 3.0 \times 10^{-4}$, whereas the reference choice $V=u_m$ gives $C'_{\rm ours} \approx 2.09 \times 10^{-4}$. The rough-wall coefficient has been used in operational models \citep{kubat2007,keghouche2010,gladstone2001,eltahan1987,eik2009}, but White's original experiments evaluated only the smooth-wall form.

\textbf{(2) Velocity scaling ambiguity and calibration.} We write $V=\alpha u_m$, using the horizontal orbital velocity as a convenient representative scale. Least-squares fits give $\alpha_\mathrm{fit}=0.684$ for $T=1.54$~s and $0.612$ for $T=0.87$~s, a relative difference of approximately 11\%. The corresponding effective coefficients, $1.43\times10^{-4}$ and $1.28\times10^{-4}$, are close to White's published value. Because $\alpha$ is fitted to these same observations, this proximity is a calibration result rather than independent validation or recovery of White's undocumented reduction.

\textbf{(3) Friction coefficient specification.} The wave friction coefficient $C_f$ is not given explicitly in White~(1980). We provide the Lambert-W solution of the rough-turbulent relation of \citet{jonsson1966} and document the accuracy range of its power-law approximation (\ref{app:Cf_derivation}).

\textbf{(4) Depth-resolved formulation.} White's waterline rough-wall expression is extended to the depth-resolved melt rate $V_m(z)$ (Eq.~\eqref{eq:Vm_full}) under finite-depth linear wave kinematics.

\textbf{(5) Splash-zone behaviour.} Experimental profiles show the melt-rate peak located below mean water level rather than at it---a behaviour not captured by the standard White~(1980) formulation. A simple contact-fraction diagnostic helps explain the observed rolloff toward the waterline, but it does not replace the need for a dedicated treatment of run-up, aeration, impact, and turbulent heat transfer in the splash zone.

Future work should target three areas: (i) experiments with measured near-ice temperature profiles, measured roughness, and wave measurements at the ice face; (ii) nonlinear and breaking-wave extensions for run-up, aeration, and turbulent transport; and (iii) independent laboratory and field tests across broader forcing ranges. The present results provide a transparent reconstruction and a two-condition laboratory calibration of White's rough-wall model, while leaving field transfer and regime independence to be established.


\appendix

\section{Rough-turbulent wave friction coefficient}
\label{app:Cf_derivation}

The rough-turbulent wave friction coefficient $C_f$ appearing in Eq.~\eqref{eq:Cf} is not given explicitly in White (1980). The power-law expression in Eq.~\eqref{eq:Cf} was obtained by fitting the rough-turbulent wave-friction relation of \citet{jonsson1966}.

Jonsson's Eq.~5.17 implicitly defines the rough-turbulent wave friction factor $f_w$, denoted $C_f$ in the present paper, through
\begin{equation}
\frac{1}{4\sqrt{f_w}} + \log_{10}\!\left(\frac{1}{4\sqrt{f_w}}\right)
= -0.08 + \log_{10}\!\left(\frac{a_{1m}}{k_s}\right).
\label{eq:Jonsson_517}
\end{equation}
Equation~\eqref{eq:Jonsson_517} is of Lambert-W type and admits the closed-form expression
\begin{equation}
C_f = \left(\frac{\ln 10}{4\,W\!\left(1.916\,\dfrac{a_{1m}}{k_s}\right)}\right)^{\!2},
\label{eq:Cf_LambertW}
\end{equation}
where $W(\cdot)$ denotes the principal branch of the Lambert-W function. The numerical constant is
\[
1.916 = 10^{-0.08}\ln 10,
\]
as obtained directly from Eq.~\eqref{eq:Jonsson_517}.

\begin{figure}[htbp]
    \centering
    \includegraphics[width=0.82\textwidth]{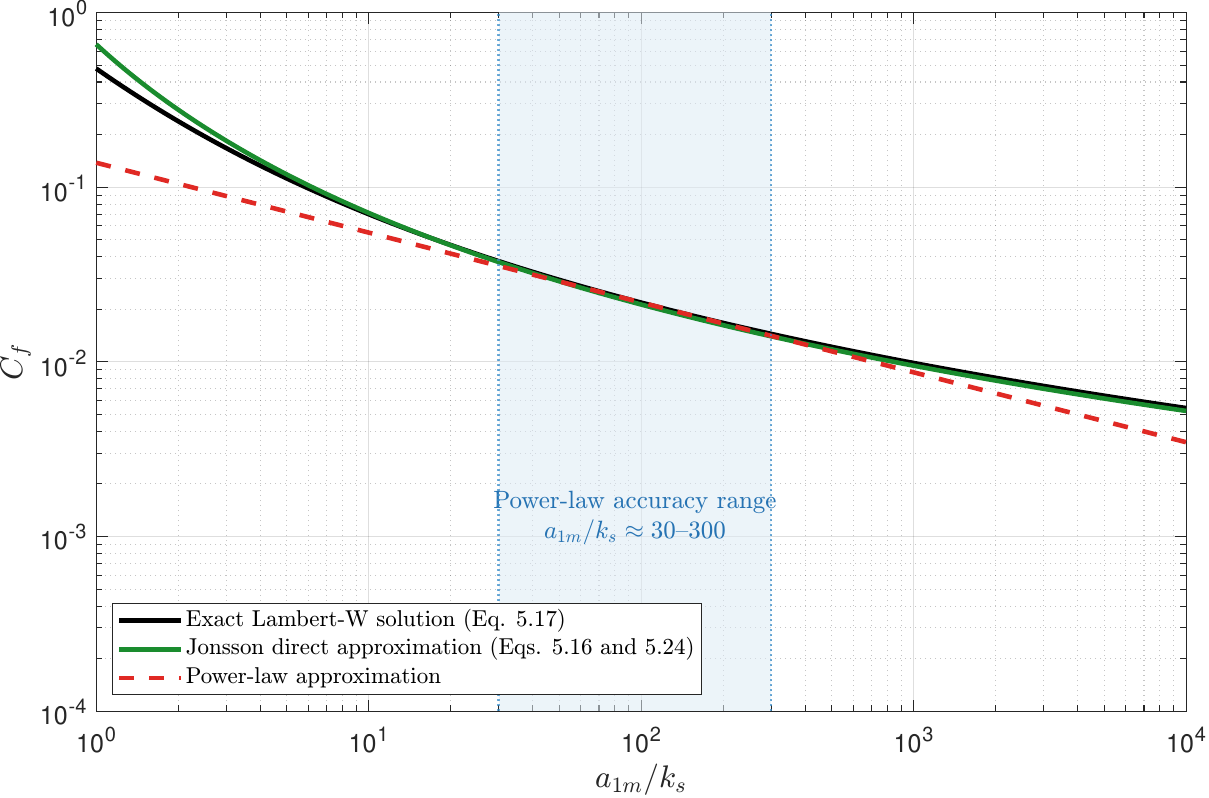}
    \caption{Comparison of wave friction coefficient $C_f$ as a function of relative roughness
             $a_{1m}/k_s$. Solid black line: exact Lambert-W solution of Eq.~\ref{eq:Jonsson_517};
             green solid line: Jonsson's (1966) direct approximation (Eqs.~5.16 and 5.24);
             red dashed line: power-law fit $C_f = 0.138\,(a_{1m}/k_s)^{-0.4}$ (Eq.~\ref{eq:Cf}).
             The shaded band indicates the power-law accuracy range $a_{1m}/k_s \approx 30$--$300$;
             the laboratory conditions
             ($a_{1m}/k_s \approx 4$--$6$) lie below this range.}
    \label{fig:Cf_comparison}
\end{figure}

Jonsson's Eq.~5.16 separately determines the boundary-layer thickness $\delta$,
\begin{equation}
\left(\frac{30\delta}{k_s}\right)\log_{10}\!\left(\frac{30\delta}{k_s}\right)
= 1.2\,\frac{a_{1m}}{k_s},
\label{eq:Jonsson_516}
\end{equation}
and is used with his Eq.~5.24 to form the alternative direct approximation shown in Fig.~\ref{fig:Cf_comparison}. It is not required for the exact inversion of Eq.~\eqref{eq:Jonsson_517}.

Evaluating Eq.~\eqref{eq:Cf_LambertW} over the range $a_{1m}/k_s \in [1,\,10^4]$ and fitting a power law by least-squares regression yields
\[
C_f = 0.138\left(\frac{a_{1m}}{k_s}\right)^{-0.4},
\]
which reproduces Eq.~\eqref{eq:Cf_LambertW} to within approximately 7\% over $a_{1m}/k_s \approx 30$--$300$ (Fig.~\ref{fig:Cf_comparison}). Jonsson's approximation based on Eqs.~5.16 and~5.24 is also shown for reference.

For the laboratory conditions in this study, $a_{1m}/k_s \approx 4$--$6$, below the accuracy range of the power-law approximation.

\section{Compact Stanton-number derivation}
\label{app:St_algebra}

\subsection*{B.1 Starting equations from Jonsson (1966) and White (1974)}

White (1980) relates the Stanton number to wave-induced boundary friction via the scaling relation
\begin{equation}
\mathrm{St} = \mathrm{St}_0 \sqrt{\frac{C_f}{C_{f0}}},
\label{eq:St_scaling}
\end{equation}
where $\mathrm{St}_0$ and $C_{f0}$ are reference values for a hydraulically smooth boundary and $C_f$ is the local rough-turbulent wave friction coefficient at the ice surface.

For a smooth boundary in the turbulent oscillatory regime, the reference friction coefficient is
\begin{equation}
C_{f0} = 0.09\,\mathrm{Re}_a^{-0.2},
\label{eq:Cf0}
\end{equation}
where the amplitude Reynolds number is
\begin{equation}
\mathrm{Re}_a = \frac{u_m a_{1m}}{\nu},
\label{eq:Rea}
\end{equation}
with $u_m$ the maximum horizontal orbital velocity, $a_{1m}$ the orbital excursion amplitude, and $\nu$ the kinematic viscosity of water.

The smooth-wall reference Stanton number follows from White's turbulent heat-transfer analogy \citep[p.~564]{white1974}:
\begin{equation}
\mathrm{St}_0
=
\frac{\tfrac{1}{2}C_{f0}}
{1 + 12.8\,(\mathrm{Pr}^{0.68}-1)\,\sqrt{\tfrac{1}{2}C_{f0}}},
\label{eq:St0}
\end{equation}
where $\mathrm{Pr}$ is the Prandtl number.

\subsection*{B.2 Derivation of the compact Stanton-number form}

Starting from Eqs.~\eqref{eq:St_scaling}--\eqref{eq:St0} and the power-law rough-wall friction coefficient (Eq.~\eqref{eq:Cf}), the compact expression for $\mathrm{St}$ (Eq.~\eqref{eq:StWhite}) is derived as follows.

\paragraph{Prandtl-number sensitivity.}
For the reference value $\mathrm{Pr} \approx 13$ used in the model, the Prandtl-number correction
\begin{equation}
\mathcal{B}(\mathrm{Pr}) = \frac{3.84}{\sqrt{2}}\,\left(\mathrm{Pr}^{0.68}-1\right) \approx 12.8
\label{eq:B_Pr}
\end{equation}
dominates the Stanton-number denominator across both laboratory ($\mathrm{Re}_a \sim 10^3$--$10^4$, $\mathrm{Re}_a^{0.1} \approx 2$--$2.5$) and field ($\mathrm{Re}_a \sim 10^5$--$10^7$, $\mathrm{Re}_a^{0.1} \approx 3$--$4$) conditions. The amplitude Reynolds number therefore contributes weakly, and the melt-rate profile is primarily controlled by the kinematic depth-dependence and the calibration coefficient $\alpha$.

\paragraph{Algebraic reduction.}
Substituting Eqs.~\eqref{eq:St0} and~\eqref{eq:Cf0} into Eq.~\eqref{eq:St_scaling}, using
$\sqrt{C_{f0}} = 0.3\,\mathrm{Re}_a^{-0.1}$ from Eq.~\eqref{eq:Cf0}, and multiplying numerator
and denominator by $\mathrm{Re}_a^{0.1}$,
\begin{equation}
\mathrm{St}
=
\frac{0.15\,\sqrt{C_f}}
{\mathrm{Re}_a^{0.1} + \dfrac{3.84}{\sqrt{2}}\,(\mathrm{Pr}^{0.68}-1)}.
\label{eq:A_St_step2}
\end{equation}
Using Eq.~\eqref{eq:Cf}, the required square root and numerical product are
\[
\sqrt{C_f}=0.3715\left(\frac{k_s}{a_{1m}}\right)^{0.2},
\qquad 0.15\times0.3715=0.0557.
\]
Equation~\eqref{eq:A_St_step2} therefore becomes
\begin{equation}
\mathrm{St}
=
\frac{0.0557\,\left(\dfrac{k_s}{a_{1m}}\right)^{0.2}}
{\mathrm{Re}_a^{0.1} + \dfrac{3.84}{\sqrt{2}}\,(\mathrm{Pr}^{0.68}-1)},
\label{eq:A_StWhite_derived}
\end{equation}
which reproduces Eq.~\eqref{eq:StWhite} in the main text.

\section{Mathematical discrepancy in the White (1980) waterline formula}
\label{app:discrepancy}

Combining the melt balance and heat-flux closure, Eqs.~\eqref{eq:Vm} and~\eqref{eq:q_w}, gives
\begin{equation}
\frac{V_m T}{H}
=
\left(\frac{\rho_w c_p}{\rho_i \Gamma}\right)
\mathrm{St}
\left(\frac{VT}{H}\right)
\Delta T,
\label{eq:A_melt_general}
\end{equation}
the published White (1980) waterline expression (Eq.~\eqref{eq:White_waterline}) is recovered by evaluating all quantities at the waterline under deep-water conditions with direct wave incidence:
\begin{equation}
z = 0, \qquad kd \to \infty, \qquad \phi = 0.
\label{eq:A_deepwater_conditions}
\end{equation}

Under Eq.~\eqref{eq:A_deepwater_conditions}, the linear-wave kinematic expressions (Eqs.~\eqref{eq:um}--\eqref{eq:a1m}) reduce to
\begin{equation}
u_m(0) = \frac{\pi H}{T}, \qquad
w_m(0) = \frac{\pi H}{T}, \qquad
a_{1m}(0) = \frac{H}{2}.
\label{eq:A_waterline_kinematics}
\end{equation}

Using White's resultant-velocity definition
\begin{equation}
V^2 = u_m^2 + w_m^2,
\label{eq:Vdef}
\end{equation}
corresponding to Eq.~(69) in White (1980),
\begin{equation}
V(0)
= \sqrt{u_m(0)^2 + w_m(0)^2}
= \sqrt{2}\,\frac{\pi H}{T},
\qquad
\frac{V(0)T}{H} = \sqrt{2}\,\pi.
\label{eq:A_V_waterline}
\end{equation}

The amplitude Reynolds number at the waterline is
\begin{equation}
\mathrm{Re}_a(0)
=
\frac{u_m(0)\,a_{1m}(0)}{\nu}
=
\frac{\pi H^2}{2T\nu}.
\label{eq:A_Rea_waterline}
\end{equation}

Substituting Eqs.~\eqref{eq:A_waterline_kinematics} and~\eqref{eq:A_V_waterline} into the Stanton-number expression (Eq.~\eqref{eq:A_StWhite_derived}) with $a_{1m}(0) = H/2$,
\begin{equation}
\mathrm{St}(0)
=
\frac{0.0557\,\left(\dfrac{2k_s}{H}\right)^{0.2}}
{\mathrm{Re}_a(0)^{0.1} + \dfrac{3.84}{\sqrt{2}}\,(\mathrm{Pr}^{0.68}-1)}.
\label{eq:A_St_waterline}
\end{equation}

Since $\mathcal{B}(\mathrm{Pr}) = (3.84/\sqrt{2})\,(\mathrm{Pr}^{0.68}-1)$ is nearly constant for a fixed fluid and dominates $\mathrm{Re}_a^{0.1}$ over the oceanographically relevant range of wave conditions, Eq.~\eqref{eq:A_St_waterline} is approximated as
\begin{equation}
\mathrm{St}(0)
\approx
\frac{0.0557\,\cdot 2^{0.2}}{\mathcal{B}(\mathrm{Pr})}
\left(\frac{k_s}{H}\right)^{0.2}.
\label{eq:A_St_approx}
\end{equation}

Inserting Eqs.~\eqref{eq:A_V_waterline} and~\eqref{eq:A_St_approx} into Eq.~\eqref{eq:A_melt_general} yields the compact waterline form
\begin{equation}
\frac{V_m T}{H}
\approx
C'\left(\frac{k_s}{H}\right)^{0.2}\Delta T,
\label{eq:A_waterline_result}
\end{equation}
where the dimensional coefficient is
\begin{equation}
C'
=
\left(\frac{\rho_w c_p}{\rho_i \Gamma}\right)
(\sqrt{2}\,\pi)
\cdot
\frac{0.0557\,\cdot 2^{0.2}}{\mathcal{B}(\mathrm{Pr})}.
\label{eq:A_Cprime}
\end{equation}

Evaluating with the physical constants listed in Table~\ref{tab:ice_properties} gives $C' \approx 3.0\times10^{-4}$, approximately twice the coefficient $1.462\times10^{-4}$ in the published White~(1980) waterline formula (Eq.~\eqref{eq:White_waterline}). This appendix follows White's own velocity definition $V = \sqrt{u_m^2 + w_m^2}$; the factor $\sqrt{2}\,\pi$ in Eq.~\eqref{eq:A_Cprime} reflects this choice.

One likely explanation for the additional factor-of-two discrepancy between $3.0\times10^{-4}$ and the published $1.46\times10^{-4}$ is that White calibrated his compact waterline formula against his own laboratory experiments, which involved comparing predicted melt profiles against the smooth-wall Eq.~(77a) of the original report rather than the rough-wall formulation. White's experiments used ice-block mass loss and waterline notch depth as the sole observables under smooth-surface conditions. The rough-wall coefficient $1.46\times10^{-4}$ was therefore not independently evaluated against experimental data in the original study; it follows from the rough-wall boundary-layer closure of \citet{jonsson1966}.

The relevant baseline for the present study is not $C' \approx 3.0\times10^{-4}$ but rather $C'_{\rm ours} = C'/\sqrt{2} \approx 2.09\times10^{-4}$, which corresponds to the reference velocity choice $V = u_m$ ($\alpha = 1$) embedded in Eq.~\eqref{eq:Vm_full} via the prefactor $0.0027 = C_{\rm thermo}\,\pi\cdot 0.0557\cdot 2^{0.2}$. The present study adopts $C'_{\rm ours}$ as the theoretical baseline and treats the remaining discrepancy with White's published value as a source of systematic uncertainty to be quantified by laboratory calibration (Section~\ref{sec:results_comparison}).

\section*{Funding}
This work received internal financial support from the Norwegian University of Science and Technology (NTNU), the Norwegian Polar Institute (NPI), and Universit\'{e} de Caen Normandie. The work was also supported by the Coastal and Marine Engineering and Management (CoMEM) Erasmus Mundus programme, funded by the European Union, with additional financial support from Arctic Integrated Solutions (ArcISo).

\section*{CRediT authorship contribution statement}
Wenjun Lu: Conceptualization, Methodology, Software, Formal analysis, Investigation, Data curation, Visualization, Writing -- original draft, Writing -- review and editing. Behnam Ghadimi: Methodology, Investigation, Writing -- review and editing. Dominique Mouaze: Methodology, Investigation, Writing -- review and editing. Marianne Font: Methodology, Investigation, Writing -- review and editing. R\'{e}mi Lambert: Methodology, Investigation, Writing -- review and editing. Harvey Goodwin: Conceptualization, Writing -- review and editing. Widar Weizhi Wang: Conceptualization, Writing -- review and editing. Raed Lubbad: Conceptualization, Writing -- review and editing. Sveinung L\o set: Conceptualization, Writing -- review and editing.

\section*{Declaration of competing interest}
The authors declare that they have no known competing financial interests or personal relationships that could have appeared to influence the work reported in this paper.

\section*{Data availability}
The experimental data comprise legacy laboratory records that have not yet been deposited in a public repository. The processed data and MATLAB analysis code supporting this study are available from the corresponding author upon reasonable request.

\section*{Declaration of generative AI and AI-assisted technologies in the manuscript preparation process}
During the preparation of this work, the authors used LLMs to support language editing, manuscript organization, consistency checks, LaTeX preparation and coding assistance. After using these tools and services, the authors reviewed and edited the content as needed and take full responsibility for the content of the published article.

\bibliographystyle{elsarticle-harv}
\bibliography{references}

\begin{thebibliography}{15}
\expandafter\ifx\csname natexlab\endcsname\relax\def\natexlab#1{#1}\fi
\providecommand{\url}[1]{\texttt{#1}}
\providecommand{\href}[2]{#2}
\providecommand{\path}[1]{#1}
\providecommand{\DOIprefix}{doi:}
\providecommand{\ArXivprefix}{arXiv:}
\providecommand{\URLprefix}{URL: }
\providecommand{\Pubmedprefix}{pmid:}
\providecommand{\doi}[1]{\href{http://dx.doi.org/#1}{\path{#1}}}
\providecommand{\Pubmed}[1]{\href{pmid:#1}{\path{#1}}}
\providecommand{\bibinfo}[2]{#2}
\ifx\xfnm\relax \def\xfnm[#1]{\unskip,\space#1}\fi
\bibitem[{Barnhart et~al.(2014)Barnhart, Anderson, Overeem, Wobus, Clow and
  Urban}]{barnhart2014}
\bibinfo{author}{Barnhart, K.R.}, \bibinfo{author}{Anderson, R.S.},
  \bibinfo{author}{Overeem, I.}, \bibinfo{author}{Wobus, C.},
  \bibinfo{author}{Clow, G.D.}, \bibinfo{author}{Urban, F.E.},
  \bibinfo{year}{2014}.
\newblock \bibinfo{title}{Modeling erosion of ice-rich permafrost bluffs along
  the {Alaskan} {Beaufort Sea} coast}.
\newblock \bibinfo{journal}{Journal of Geophysical Research: Earth Surface}
  \bibinfo{volume}{119}, \bibinfo{pages}{1155--1179}.
\newblock \DOIprefix\doi{10.1002/2013JF002845}.
\bibitem[{Eik(2009)}]{eik2009}
\bibinfo{author}{Eik, K.}, \bibinfo{year}{2009}.
\newblock \bibinfo{title}{Iceberg deterioration in the {Barents Sea}}, in:
  \bibinfo{booktitle}{Proceedings of the International Conference on Port and
  Ocean Engineering Under Arctic Conditions}, pp. \bibinfo{pages}{POAC09--113}.
\bibitem[{El-Tahan et~al.(1987)El-Tahan, Venkatesh and El-Tahan}]{eltahan1987}
\bibinfo{author}{El-Tahan, M.}, \bibinfo{author}{Venkatesh, S.},
  \bibinfo{author}{El-Tahan, H.}, \bibinfo{year}{1987}.
\newblock \bibinfo{title}{Validation and quantitative assessment of the
  deterioration mechanisms of {Arctic} icebergs}.
\newblock \bibinfo{journal}{Journal of Offshore Mechanics and Arctic
  Engineering} \bibinfo{volume}{109}, \bibinfo{pages}{102--108}.
\newblock \DOIprefix\doi{10.1115/1.3256981}.
\bibitem[{Forouzi~Feshalami et~al.(2025a)Forouzi~Feshalami, L{\o}set, Lubbad,
  Lu, Skourup and Kashafutdinov}]{forouzi2025a}
\bibinfo{author}{Forouzi~Feshalami, B.}, \bibinfo{author}{L{\o}set, S.},
  \bibinfo{author}{Lubbad, R.}, \bibinfo{author}{Lu, W.},
  \bibinfo{author}{Skourup, H.}, \bibinfo{author}{Kashafutdinov, M.},
  \bibinfo{year}{2025}a.
\newblock \bibinfo{title}{A numerical model for the simulation of wave-induced
  erosion of floating icebergs: {Implementation} and validation against wave
  flume data}.
\newblock \bibinfo{journal}{Journal of Ocean Engineering and Science}
  \bibinfo{volume}{10}, \bibinfo{pages}{968--981}.
\newblock \DOIprefix\doi{10.1016/j.joes.2025.01.003}.
\bibitem[{Forouzi~Feshalami et~al.(2025b)Forouzi~Feshalami, Lu, L{\o}set,
  Lubbad, Skourup and H{\o}yland}]{forouzi2025b}
\bibinfo{author}{Forouzi~Feshalami, B.}, \bibinfo{author}{Lu, W.},
  \bibinfo{author}{L{\o}set, S.}, \bibinfo{author}{Lubbad, R.},
  \bibinfo{author}{Skourup, H.}, \bibinfo{author}{H{\o}yland, K.V.},
  \bibinfo{year}{2025}b.
\newblock \bibinfo{title}{An artificial neural network correlation for the
  wave-induced melt rate of floating icebergs in the {Barents Sea}}.
\newblock \bibinfo{journal}{Cold Regions Science and Technology} ,
  \bibinfo{pages}{104575}\DOIprefix\doi{10.1016/j.coldregions.2025.104575}.
\bibitem[{Gladstone et~al.(2001)Gladstone, Bigg and Nicholls}]{gladstone2001}
\bibinfo{author}{Gladstone, R.M.}, \bibinfo{author}{Bigg, G.R.},
  \bibinfo{author}{Nicholls, K.W.}, \bibinfo{year}{2001}.
\newblock \bibinfo{title}{Iceberg trajectory modeling and meltwater injection
  in the {Southern Ocean}}.
\newblock \bibinfo{journal}{Journal of Geophysical Research: Oceans}
  \bibinfo{volume}{106}, \bibinfo{pages}{19903--19915}.
\newblock \DOIprefix\doi{10.1029/2000JC000347}.
\bibitem[{Jonsson(1966)}]{jonsson1966}
\bibinfo{author}{Jonsson, I.G.}, \bibinfo{year}{1966}.
\newblock \bibinfo{title}{Wave boundary layers and friction factors}, in:
  \bibinfo{booktitle}{Proceedings of the 10th International Conference on
  Coastal Engineering}, \bibinfo{publisher}{ASCE}. pp.
  \bibinfo{pages}{127--148}.
\newblock \DOIprefix\doi{10.9753/icce.v10.9}.
\bibitem[{Keghouche et~al.(2010)Keghouche, Counillon and
  Bertino}]{keghouche2010}
\bibinfo{author}{Keghouche, I.}, \bibinfo{author}{Counillon, F.},
  \bibinfo{author}{Bertino, L.}, \bibinfo{year}{2010}.
\newblock \bibinfo{title}{Modeling dynamics and thermodynamics of icebergs in
  the {Barents Sea} from 1987 to 2005}.
\newblock \bibinfo{journal}{Journal of Geophysical Research: Oceans}
  \bibinfo{volume}{115}.
\newblock \DOIprefix\doi{10.1029/2010JC006165}.
\bibitem[{Kubat et~al.(2007)Kubat, Sayed, Savage, Carrieres and
  Crocker}]{kubat2007}
\bibinfo{author}{Kubat, I.}, \bibinfo{author}{Sayed, M.},
  \bibinfo{author}{Savage, S.B.}, \bibinfo{author}{Carrieres, T.},
  \bibinfo{author}{Crocker, G.}, \bibinfo{year}{2007}.
\newblock \bibinfo{title}{An operational iceberg deterioration model}, in:
  \bibinfo{booktitle}{Proceedings of the Seventeenth International Offshore and
  Polar Engineering Conference}, \bibinfo{publisher}{ISOPE}. pp.
  \bibinfo{pages}{ISOPE--I--07--282}.
\bibitem[{Mansard and Funke(1980)}]{mansardfunke1980}
\bibinfo{author}{Mansard, E.P.D.}, \bibinfo{author}{Funke, E.R.},
  \bibinfo{year}{1980}.
\newblock \bibinfo{title}{The measurement of incident and reflected spectra
  using a least squares method}, in: \bibinfo{booktitle}{Proceedings of the
  17th International Conference on Coastal Engineering},
  \bibinfo{publisher}{ASCE}, \bibinfo{address}{Sydney, Australia}. pp.
  \bibinfo{pages}{154--172}.
\newblock \DOIprefix\doi{10.9753/icce.v17.8}.
\bibitem[{R{\"o}hl(2006)}]{rohl2006}
\bibinfo{author}{R{\"o}hl, K.}, \bibinfo{year}{2006}.
\newblock \bibinfo{title}{Thermo-erosional notch development at
  fresh-water-calving {Tasman Glacier}, {New Zealand}}.
\newblock \bibinfo{journal}{Journal of Glaciology} \bibinfo{volume}{52},
  \bibinfo{pages}{203--213}.
\newblock \DOIprefix\doi{10.3189/172756506781828773}.
\bibitem[{Sartore et~al.(2025)Sartore, Wagner, Siegfried, Pujara and
  Zoet}]{sartore2025}
\bibinfo{author}{Sartore, N.B.}, \bibinfo{author}{Wagner, T.J.W.},
  \bibinfo{author}{Siegfried, M.R.}, \bibinfo{author}{Pujara, N.},
  \bibinfo{author}{Zoet, L.K.}, \bibinfo{year}{2025}.
\newblock \bibinfo{title}{Wave erosion, frontal bending, and calving at {Ross
  Ice Shelf}}.
\newblock \bibinfo{journal}{The Cryosphere} \bibinfo{volume}{19},
  \bibinfo{pages}{249--265}.
\newblock \DOIprefix\doi{10.5194/tc-19-249-2025}.
\bibitem[{White(1974)}]{white1974}
\bibinfo{author}{White, F.M.}, \bibinfo{year}{1974}.
\newblock \bibinfo{title}{Viscous Fluid Flow}.
\newblock \bibinfo{publisher}{McGraw-Hill}, \bibinfo{address}{New York}.
\bibitem[{White et~al.(1980)White, Spaulding and Gominho}]{white1980}
\bibinfo{author}{White, F.M.}, \bibinfo{author}{Spaulding, M.L.},
  \bibinfo{author}{Gominho, L.}, \bibinfo{year}{1980}.
\newblock \bibinfo{title}{Theoretical Estimates of the Various Mechanisms
  Involved in Iceberg Deterioration in the Open Ocean}.
\newblock \bibinfo{type}{Technical Report} \bibinfo{number}{CG-D-62-80}. United
  States Coast Guard, Research and Development Center.
  \bibinfo{address}{Groton, CT, USA}.
\bibitem[{Wolterman et~al.(2026)Wolterman, Wagner, Zoet and
  Pujara}]{wolterman2026}
\bibinfo{author}{Wolterman, A.}, \bibinfo{author}{Wagner, T.J.W.},
  \bibinfo{author}{Zoet, L.K.}, \bibinfo{author}{Pujara, N.},
  \bibinfo{year}{2026}.
\newblock \bibinfo{title}{Wave erosion of ice cliffs: melt rate due to
  reflection of non-breaking surface waves}.
\newblock \bibinfo{journal}{Journal of Fluid Mechanics} \bibinfo{volume}{1036},
  \bibinfo{pages}{A29}.
\newblock \DOIprefix\doi{10.1017/jfm.2026.11603}.

\end{thebibliography}

\end{document}